\documentclass[11pt,a4paper]{article}
\pdfoutput=1
\usepackage{jheppub}
\usepackage{adjustbox}

\usepackage{amsmath,amssymb,amsbsy,amsfonts,latexsym,graphicx}
\usepackage{hyperref}
\usepackage{float}
\usepackage{color, array, subfigure}

\newcommand{\Tr}{{\rm Tr}\,}

\allowdisplaybreaks

\title{  Dynamical Compactification with Matter}
\author[a]{Kyung Kiu Kim,}
\author[b,c]{Seoktae Koh}
\author[b]{Gansukh Tumurtushaa}
 
\affiliation[a]{College of General Education, Kookmin University, Seoul 02707, Korea}
\affiliation[b]{Department of Science Education, Jeju National University, Jeju, 63243, Korea}
\affiliation[c]{Institute for Gravitation and the Cosmos, The Pennsylvania State University, University Park, PA 16802, USA}

\emailAdd{kimkyungkiu@kookmin.ac.kr}
\emailAdd{kundol.koh@jejunu.ac.kr}
\emailAdd{gansukh@jejunu.ac.kr}

\abstract{In this work, we study cosmological solutions of the 8--dimensional Einstein Yang-Mills theory coupled to a perfect-fluid matter. A Yang-Mills instanton of extra dimensions causes a 4--dimensional expanding universe with dynamical compactification of the extra dimensions. To construct physically reliable situations, we impose the null energy condition on the matter. This energy condition is affected by the extra dimensions. Then, we consider cosmological constant to grasp the structure of the solution space. Even in this simple case, we find several interesting solutions, such as bouncing universes and oscillatory solutions, eventually arriving at a de Sitter universe with stabilized compact dimensions. In addition, we consider a class of matters whose energy density depends on the volume of the extra dimensions. This case shows another set of bouncing universes. Also, a real scalar with potential is taken into account. The scalar field model admits de Sitter solutions due to the choice of potential, and we demonstrate how potentials can be constructed using flow equations. {\color{black}Thus, what we discuss in this work is based on the 8--dimensional Einstein frame, which corresponds to the 4--dimensional Jordan frame by dimensional reduction. Consequently, the results are derived in the 4--dimensional Jordan frame, not in the 4--dimensional Einstein frame.}}


\begin{document}
\maketitle
\flushbottom

\section{Introduction}

General relativity is one of the most successful theories in physics. Numerous predictions have been suggested based on this theory, and various observations and experiments have confirmed such predictions. Although this theory is the only proven theory to describe our universe, it has a fatal weakness that makes a quantum version impossible. To remedy this weakness, many extensions of general relativity have been proposed. One of the promising extended theories is string theory. String theory has several consequences. The first notable consequence is that our spacetime is more than 4--dimensions, and the second is that not only gravity but also gauge theory plays an important role  \cite{Polchinski:1998rq, Green:1987sp}. 

Many promising cosmological models employ extra dimensions~\cite{Arkani-Hamed:1998sfv, Csaki:1999mp} and lead us to accept the speculation that some gauge theory plays an important role in the extra dimensions. Since we cannot observe the extra dimensions, one of the possible construction of our universe is to render the extra dimensions compact and make their size sufficiently small. In such compact manifolds, most of the long-lived effects from gauge theory are topological objects. Accordingly, an instanton solution can be regarded as a major physical object in the extra dimensions whose signature is Euclidean. The extra dimensions of a topological object are also worth considering as models in cosmology.  
On the other hand, the smallness of extra dimensions may also lead to another hierarchy problem. It is difficult to find a reason why our universe begins with large 4--dimensions accompanied by extremely small extra dimensions. Even though there are some studies about this question using matrix theories \cite{Kim:2011ts, Kim:2012mw, Aoki:2019tby, Nishimura:2020blu}, it has not been well addressed in the gravity theory context. 

Besides, our universe is expanding. The expanding behavior may be related to the extra-dimension dynamics, approaching the small size. This possibility is explored in a toy model based on the 8--dimensional Yang-Mills (YM) theory \cite{Kim:2018mfv}. Relating the expanding universe to a compactification of extra dimensions is not something new. A number of studies have already explored this idea in different ways \cite{Shafi:1983hj, Randjbar-Daemi:1983awk, Kolb:1984ga, Okada:1984cv, Maeda:1985ka, Accetta:1986vq, Kolb:1990vq, Mazumdar:1999tk, Steinhardt:2008nk}. Also, similar considerations of topological objects in extra dimensions have been discussed in \cite{Horvath:1978dg, Randjbar-Daemi:1982opc, Salam:1984cj, Randjbar-Daemi:1983zks, Randjbar-Daemi:1983xth, Kihara:2009ea, OColgain:2009qab}. However, there is no clear explanation for the small extra dimensions.  
Thus, in this paper, we will continue and extend the discussion of \cite{Kim:2018mfv} through the interplay between expanding universe and dynamical extra dimensions by considering matters. Such an extended model can be a nice toy model to see the role of matter in the evolution of the universe. We take the perfect fluid into account as an appropriate matter. 

To avoid unphysical situations in our model, the null energy condition (NEC) will be used. We derive the NEC for our model and find that the condition has a contribution from the extra dimensions. In our parallel work \cite{KKYH001}, we discuss the Maldacena--Nunez no--go theorem without matter to understand why an expanding universe needs dynamical compactification in the context of our model. As a key observable of the theorem, the scalar curvature of the 4--dimensional universe is taken into account. Thus, we find the effects of matter on the scalar curvature explaining various dynamical compactifications. 
In the present study, we consider three kinds of perfect fluid matters coupled with the Einstein--YM theory. First, we consider the cosmological constant to get an insight into the analysis of the matter effect. Then, the energy density that depends on the volume of the extra dimensions is discussed in the second case. Lastly, a real scalar field is introduced. 

The case with cosmological constant has revealed various interesting cosmological evolutions such as bouncing solutions and de Sitter solutions. The full solution space can be analyzed completely by considering two--dimensional vector fields dubbed `solution flows' like those in \cite{Kim:2018mfv}. The bouncing solution shows a bouncing behavior of the Hubble parameter with convex bouncing extra dimension size. Therefore, the three--dimensional space in the universe and the extra dimensions have seesaw-like dynamics through the cosmological constant. In addition, we find a stable de Sitter solution described by a fixed point of dynamical equations. We discuss a cosmological constant problem in this toy model. Next, we consider a matter satisfying the NEC and a generalized matter whose energy density depends on the volume of the extra dimensions. The resulting cosmological solutions show the bouncing behavior of the universe and the extra dimensions. In the scalar field case, we focus on de Sitter solutions. Thus, we demand a constant Hubble parameter. This requirement enables us to find the evolution of the extra dimensions and the scalar field potential. The scalar potential has a nontrivial profile due to the extra dimension dynamics. 

The main goal of this paper is to understand the role of matter in the accelerating universe and the dynamical compactification of the extra dimensions qualitatively. Thus, it is not in the scope of our work to relate our results of the present work to observations in cosmology, and a phenomenological study based on this qualitative research will be reported as our future work. {\color{black} It is also notable that we adhere to the 8--dimensional formulation in this paper. Since any effective action obtained by the compactification of the extra dimensions does not guarantee a consistent truncation, one needs sometimes additional constraints to the 4--dimensional effective action. For this reason, we adopt the 8--dimensional Einstein frame in this work. For interested readers, we provide the 4--dimensional action regarded as the Jordan frame in Appendix \ref{Decom-Einstein} for the case without matter.}

This paper is organized as follows. In section 2, we present the general formulation for a perfect fluid matter and discuss the NEC and the scalar curvature of our toy model. In section 3, we consider a case where the cosmological constant is the matter and show various resultant cosmological solutions. In section 4, we consider more nontrivial matters. The first case takes a matter whose energy-momentum tensor depends on the extra dimension volume into account. The second case considers the scalar field, where we explain how to obtain the scalar field potential for a de Sitter universe. We conclude our results and provide future directions in section 5.

\section{Instanton Universe with Perfect Fluid Matter}\label{sec2}

This section provides a formulation of the 8--dimensional Einstein--YM theory coupled to a matter. The basic formulation without matter was provided by \cite{Kim:2018mfv}. 

\subsection{Einstein Yang-Mills with Instanton}\label{sec2.1}

We start with the following action:
\begin{align}
S=\int d^8 x \sqrt{-G} \left( \frac{1}{16\pi G_{(8)}} R + \frac{1}{4 g_{\text{YM}}^2} \Tr F_{MN}F^{MN} + \mathcal{L}_m \right)~,
\end{align}
where $G_{(8)}$ and $g_{\text{YM}}$ are the 8--dimensional Newton constant and the YM coupling, respectively. $\mathcal{L}_m$ denotes a matter action that does not couple to the YM field directly. In this work, we will consider $SU(2)$ as the gauge group of the YM fields. Then, the equations of motion are as follows:
\begin{align}
&R_{MN} - \frac{1}{2} G_{MN} R - 8 \pi G_{(8)} T^{\text{YM}}_{MN} = 8 \pi G_{(8)} T^m_{MN}\,,\label{eomEin}\\
&D_M F^{MN} = 0\,,\label{eomYM}
\end{align}
where $T^{\text{YM}}_{MN}$ and $T^m_{MN}$ denote the energy-momentum tensors from the YM field and the matter described by $\mathcal{L}_m$. The explicit form of $T^{\text{YM}}_{MN}$ is given by
\begin{align}
T_{MN}^{\text{YM}} = -\frac{1}{g_{YM}^2} \Tr \left( G^{PQ} F_{MP} F_{NQ} - \frac{1}{4}G_{MN}F_{PQ}F^{PQ} \right)~.
\end{align} 
Also, $D_M$ represents the gauge covariant derivative defined by
\begin{align}
D_M \equiv \nabla_M + \left[ A_M,~~~\right]~,
\end{align}
where $\nabla_M$ is the gravity covariant derivative. The gauge field and the field strength are matrix-valued as follows:
\begin{align}
A_M =\sum_{i=1}^3 A_M^i \tau^i ~,~ F_{MN} = \partial_M A_N - \partial_N A_M + \left[A_M,A_N \right]~.
\end{align} 
Here $\tau^i$'s are the generator of the $SU(2)$ group and they satisfy the corresponding Lie algebra\footnote{The structure constant is $f^{ijk}=-\epsilon^{ijk}$ and $\tau^i= \frac{\sigma^i}{2i}$, where $\sigma^i$'s are the Pauli matrices.}:
\begin{align}
\left[ \tau^i , \tau^j\right] = f^{ijk} \tau^k~.
\end{align}
In addition, the generators are normalized by $\Tr \tau^i\tau^j= - \delta^{ij}$.

Now, we choose an ansatz for cosmological solutions. The 8--dimensional metric is restricted to the following form:
\begin{align}\label{metric00}
ds^2 = G_{MN}dX^M dX^N = g_{\mu\nu}(x) dx^\mu dx^\nu + e^{2f(x)}h_{ab}(y) dy^a dy^b~.
\end{align} 
We use coordinate decomposition given as $X^M = \left( x^\mu , y^a \right)$. The indices $M$, $\mu$, and $a$ are running from 0 to 7, from 0 to 3, and from 1 to 4, respectively. We call the spacetime described by $x^\mu$ and the space in terms of $y^a$, the (toy) universe, and the extra dimensions for convenience. A general form of the Einstein equation based on this metric is presented in Appendix \ref{Decom-Einstein}.

For the gauge field, we will take an (anti)instanton solution in the sub-manifold of the extra dimensions, whose metric is given by $ds_{h}^2=h_{ab}dy^a dy^b$. Thus the gauge one-form is
\begin{align}
A = A_a (y) dy^a~,
\end{align} 
where $A_a (y)$ is the one-(anti)instanton solution. The field strength is (anti)self-dual. This choice simplifies the equations of motion. For instance, (\ref{eomYM}) is easily satisfied by the Bianchi identity. In addition, some components of the energy-momentum tensor vanish as $T^{\text{YM}}_{ab}=T^{\text{YM}}_{\mu a}=0$. On the other hand, the nontrivial components are given by 
\begin{align}
T^{\text{YM}}_{\mu\nu} = - \frac{1}{g_{YM}^2} e^{-4 f(x)} g_{\mu\nu} \rho_1(y)~, 
\end{align}
where $\rho_1(y)$ is the (anti)instanton action density as follows:
\begin{align}
\rho_1 (y) \equiv -\frac{1}{4} h^{ab}h^{cd}\,\Tr F_{ac}F_{bd}~.
\end{align}
To find cosmological solutions, we grant homogeneity and isotropy to the universe. A conformally flat geometry will be taken for the extra dimensions due to the conformal nature of the (anti)instanton solution. Therefore, the metric ansatz becomes
\begin{align}
ds^2 = -dt^2 + e^{2h(t)} d\textbf{x}\cdot d\textbf{x}+ e^{2f(t)} e^{2 w(y)} d\textbf{y}\cdot d\textbf{y} \,,
\end{align}
where $\textbf{x}$ denotes the spatial component of $x^\mu$ and $\textbf{y}$ is nothing but $y^a$, and the inner product ``$\cdot$" is defined as a flat space inner product. As an explicit form of gauge field configuration, the action density is given by
\begin{align}
\rho_1 (y) = e^{-4w(y)} \frac{48 \zeta_s^4}{\left( |\textbf{y}-\textbf{y}_0|^2 +\zeta_s^2 \right)^4}\,,
\end{align}
where $\textbf{y}_0$ is an (anti)instanton position parameter which will be taken as the origin of the sub-manifold in the extra dimensions. So it is convenient to use a radial coordinate $r$ defined by $r^2= |\textbf{y}-\textbf{y}_0|^2$. Here, $\zeta_s$ is the size parameter of the instanton. Plugging this action density into the Einstein equation (\ref{EinsteinGEq}), one can notice that an easy way to satisfy the Einstein equation is to take a constant $\rho_1(y)$. Thus we can identify 
\begin{align}
h_{ab} = e^{2 w(r)} \delta_{ab}~,
\end{align}
where
\begin{align}
w(r) = \log \frac{\zeta_s\zeta_4}{r^2 + \zeta_s^2}~.
\end{align} 
Therefore, the metric becomes 
\begin{align}\label{Background_Metric}
ds^2 = - dt^2 + e^{2h(t)} d\textbf{x}\cdot d\textbf{x} + e^{2f(t)} ds_{(4)}^2
\end{align}
with the 4--dimensional sub-manifold, whose metric is
\begin{align}
 ds_{(4)}^2 = \left(  \frac{\zeta_4\zeta_s}{r^2 + \zeta_s^2}  \right)^2  d\textbf{y}\cdot d\textbf{y}=   \frac{\zeta_4^2}{\left(1+\tilde{r}^2\right)^2}  d\tilde{\textbf{y}}\cdot d\tilde{\textbf{y}}~,
\end{align}
where $\tilde{\textbf{y}}$ is the dimensionless coordinate defined by $\tilde{\textbf{y}}=\zeta_s \textbf{y}$. Then, this manifold is a 4--sphere with constant curvature $R_{(4)}=48/\zeta_4^2$ and volume $\pi^2\zeta_4^4/6$. Therefore the time-dependent volume of the extra dimensions is
\begin{align}
\mathcal{V}_{4}= \frac{\pi^2\zeta_4^4}{6}e^{4f(t)} \,.
\end{align}
We take into account $\mathbb{S}^4$ as extra dimensions in this work. This extra-dimension manifold can be extended to other compact manifolds. In \cite{KKYH001}, this extension will be studied. The possible compact manifolds with constant curvature give rise to the same dynamics by identifying the volume of the compact manifolds, and the corresponding curvature depends on $\zeta_4$.

{\color{black}Since our model is defined in 8--dimensional spacetimes, the starting point is the 8--dimensional Einstein frame in this paper. It is notable that this frame is different from the 4-dimensional Einstein frame. In other words, our analyses after compactification are focused on the 4--dimensional Jordan frame. The appendix \ref{Decom-Einstein} contains the relation between the two frames.}

\subsection{Inclusion of perfect-fluid matter}

So far, we have described the ansatz for the geometry and the gauge field. If there is no more matter other than the gauge field, i.e., $T_{MN}^m=0$, the resultant solutions are the same as those discussed in \cite{Kim:2018mfv}. 
Thus, we specify a matter on this geometry.  
As a matter in our study, we employ a perfect fluid given in the following form:
\begin{align}\label{perfect}
T^m_{MN}= \mathcal{E} u_M u_N + \mathcal{P}_{(3)} P_{MN}^{(3)} + \mathcal{P}_{(4)} P_{MN}^{(4)}~,
\end{align}
where $u_M$ is the unit timelike velocity field in this background (\ref{Background_Metric}). $P_{MN}^{(3)}$ and $P_{MN}^{(4)}$ are projections associated with the 3--dimensional space (described by the coordinates $\textbf{x}$'s) and the extra dimensions (described by coordinates $\textbf{y}$'s), respectively. $\mathcal{E}$ is the energy density, and $\mathcal{P}_{(3)}$ and $\mathcal{P}_{(4)}$ are pressures associated with the 3--dimensional space in the universe and the extra dimensions.  

Including the perfect fluid, the Einstein equation (\ref{eomEin}) becomes 
\begin{align}\label{eom1}
&\tilde{E}_1=4 \dot{f} \dot{h}+2 \dot{f}^2+\dot{h}^2+\frac{8 e^{-2 f}}{\zeta_4^2}-\frac{e^{-4 f}}{3 \zeta _0^2}-\frac{8\pi G_{(8)}\mathcal{E}}{3}=0\,,\nonumber\\
&\tilde{E}_2=4 \ddot{f}+8 \dot{f} \dot{h}+10 \dot{f}^2+2 \ddot{h}+3 \dot{h}^2 + \frac{24 e^{-2 f}}{\zeta_4^2}-\frac{e^{-4 f}}{\zeta _0^2}+8\pi G_{(8)}\mathcal{P}_{(3)}=0\,,\nonumber\\
&\tilde{E}_3=\ddot{f}+3 \dot{f} \dot{h}+2 \dot{f}^2+\ddot{h}+2 \dot{h}^2+\frac{4 e^{-2 f}}{\zeta_4^2}+\frac{8\pi G_{(8)}\mathcal{P}_{(4)}}{3}=0~,
\end{align}
where dotted functions denote time-derivatives of the functions, and $\zeta_0$ is defined by $\zeta_0^2 \equiv \zeta_4^4 g_{YM}^2/(384 \pi  G_{(8)})$. Since there are three equations for the dynamical functions $h$ and $f$, one of these equations must be a constraint. A combination of these equations, which reads $\tilde{E}_4=\dot{\tilde{E}}_1 + \left(4\dot{f}+3\dot{h} \right)\tilde{E}_1 - \dot{h} \tilde{E}_2 - 4 \dot{f} \tilde{E}_3$, should also vanish to satisfy the other Einstein equation. 

For homogeneity, the energy density and the pressures are required as functions of the time only. While we allow only time dependence for the energy density and pressures, the vanishing condition $\tilde{E}_4=0$ is
\begin{align}
\dot{\mathcal{E}}+ 3\left(\mathcal{E}+\mathcal{P}_{(3)} \right) \dot{h}+4 (\mathcal{E}+ \mathcal{P}_{(4)} )\dot{f}   = 0~.
\end{align}
This is nothing but the Bianchi identity corresponding to energy-momentum conservation. It is easy to see that one of the possible perfect fluids satisfying this conservation law is the cosmological constant. The following subsection is focused on the investigation of this case.

Now, let us take a convenient redefinition of fields and the time scaling to simplify the equations of motion (\ref{eom1}). This is achieved by taking the following transformation:
\begin{align}\label{reScale00}
t \to \frac{\zeta_4^2}{4\sqrt{3}\zeta_0}t~,~f \to f + \frac{1}{2}\log    \left(\frac{\zeta_4^2}{12\zeta_0^2} \right)~,~\left(\mathcal{E},\mathcal{P}_{(3)},\mathcal{P}_{(4)} \right) \to \frac{6\zeta_0^2}{\pi G_{(8)} \zeta_4^4}\left(\epsilon,p_3,p_4\right)\,.
\end{align} 
Then, the equations of motion and the energy-momentum conservation law becomes
\begin{align}\label{flowEq0}
&E_1=2 X-X^2 + H^2+4 \dot{f} H+2 \dot{f}^2-\frac{\epsilon }{3}=0\,,\nonumber\\
&E_2=2 X-X^2+H^2+\frac{8 }{3}\dot{f} H+\frac{10 }{3}\dot{f}^2+ \frac{4}{3} \ddot{f}+\frac{2 }{3}\dot{H}+\frac{p_3}{3} = 0\,,\nonumber\\
&E_3=X+2 H^2+3 \dot{f} H+2 \dot{f}^2+\ddot{f}+\dot{H}+\frac{p_4}{3}=0\,,\nonumber\\
&E_4=\dot{\epsilon}+3 H \left(\epsilon +p_3\right)+4 \dot{f} \left(\epsilon+p_4 \right)=0\,,
\end{align}
where the inverse size of the extra dimensions $X$ is given by $X=e^{-2f}$, and the time-dependent Hubble parameter $H$ is defined by $H\equiv\dot{h}$. 

To find $H$ and $f$, we first need to solve the constraint equation, $E_1=0$. This equation is quadratic in $X$, so we have two roots given by
\begin{align}\label{Inverse S}
e^{-2f}=X_{\pm} = 1 \mp \sqrt{1+2 \dot{f}^2+H^2+4 \dot{f} H-\frac{\epsilon }{3}}\,,
\end{align}
where $X_+$ describes the extra dimensions whose size $e^{2f}$ is larger than 1. On the other hand, the solutions with $X_-$ correspond to the geometries with the smaller size, {\it i.e.}, $e^{2f}<1$. If we recover the original volume of the extra dimensions by inverting (\ref{reScale00}), $X_+$ or $X_-$ solutions describe the case where $\sqrt{\mathcal{V}_4}$ is larger or smaller than $\pi\sqrt{\frac{2}{3}} \frac{16\pi G_{(8)}}{g_{\text{YM}}^2}$, respectively. If we replace the coupling constant with length scales given by $16\pi G_{(8)}= 2\, l_p^6$ and $g_{\text{YM}}^2=l_{\text{YM}}^4$, the 4-dimensional volume becomes
\begin{align}
\left(\mathcal{V}_4\right)_{X=1} =  \frac{8\pi^2}{3}\left(\frac{l_p}{l_{\text{YM}}} \right)^8 l_p^4\,.
\end{align}
If we naively apply the length scales based on string theory, this volume size becomes the order of the string scale. Even though the two scales, $l_p$ and $l_{\text{YM}}$, are in a similar order, the solution set associated with $e^{-2f}=X_-$ describes the extra dimensions smaller than the Planck scale volume. Therefore, this set of solutions is hard to believe as a classical description. For this reason, we abandon the $X_-$ root and set $X=X_+$\footnote{For completeness, we summarize a different slicing of solution sets which shows the dynamics near $X=1$ in the Appendix \ref{AnotherType}.}.  

Finally, we are ready to solve the dynamical equation for the general perfect-fluid matter (\ref{perfect}). Together with $X=X_+$, the flow equation for time evolution is as follows:
\begin{align}\label{dynamical00}
&\ddot{f}= 5 \dot{f} H+2 H^2+X-\left(\frac{\omega _3}{2}-\frac{\omega _4}{3}+\frac{1}{2}\right) \epsilon\nonumber\\
&\dot{H}=-8 \dot{f} H-2 \dot{f}^2-4 H^2-2 X+\left(\frac{\omega _3}{2}-\frac{2 \omega _4}{3}+\frac{1}{2}\right) \epsilon\nonumber\\
&\dot{\epsilon}=-3 H \left(1+\omega _3\right) \epsilon -4 \dot{f} \left(1+\omega _4\right) \epsilon\,,
\end{align}
where we introduce the equations of state, $\omega_3$ and $\omega_4$, with $p_3 = \omega_3 \epsilon$ and $p_4 = \omega_4 \epsilon$. Given a set of initial values $\{\, \dot{f},H,\epsilon\,\}_{t=t_i}$ and the equations of state $\omega_3(t)$ and $\omega_4(t)$, the time-dependent functions, $\{\,\dot{f}(t),H(t),e^{-2f(t)},\epsilon(t)\,\}$, can be obtained by solving (\ref{dynamical00}). 

\subsection{Null energy condition and curvature of universe}\label{sec2.3}
When discussing gravitational systems coupled to matter, an energy condition is crucial in justifying physical reality. We adopt the NEC in this work as the criterion for this justification. One may consider the weak energy condition, but some accelerating universe solutions are possible even for negative energy in our model due to extra-dimension dynamics. Thus we don't demand the positivity of energy density.

The NEC stipulates that the contraction of the energy-momentum tensor for any null vector $n^M$ must be greater than or equal to zero. That is, it can be expressed as
\begin{align}
T_{MN} n^M n^N \geq 0
\end{align} 
for any null vector $n^M$. Since the metric (\ref{Background_Metric}) contains an isotropic 3--dimensional space ($d{\textbf{x}}\cdot d\textbf{x}$) and 4--extra dimensions ($d{\textbf{y}}\cdot d\textbf{y}$), one may introduce a generic form of the null vector given by $n^M = ( 1, n^i_{(3)}, n^a_{(4)} )$. The components of the null vector satisfy $P^{(3)}_{ij} n_{(3)}^i n_{(3)}^j = \cos^2\Theta$ and $P^{(4)}_{ab}n^a_{(4)}n^b_{(4)}=\sin^2\Theta$, where $\Theta$ can be taken as an arbitrary constant and $P^{(3)}_{ij}$ and $P^{(4)}_{ab}$ are the projections appearing in (\ref{perfect}). Then the NEC inequality becomes
\begin{align}
\mathcal{E} + \mathcal{P}_{(3)} \cos^2\Theta +\left(\mathcal{P}_{(4)} + \frac{48}{g_{\text{YM}}^2 \zeta_4^4 } e^{-4f} \right) \sin^2\Theta\geq 0\,.
\end{align}
Adopting (\ref{reScale00}) again, one may write down this inequality as
\begin{align}
\epsilon + p_3 \cos^2 \Theta + \left( p_4 + 12 X^2 \right) \sin^2 \Theta \geq 0\,.
\end{align}
Therefore, we arrive at the NEC written as follows: 
\begin{align}\label{NEC}
\epsilon + p_3 \geq 0~~\text{and}~~\epsilon + p_4  + 12 X^2 \geq 0\,.
\end{align}

As a comment on this NEC (\ref{NEC}), the first condition is the same as the usual NEC in (3+1)--dimensions. The equation of state $\omega_3$ has the minimum corresponding to a cosmological constant, i.e., $\omega_3=-1$. However, the other equation of state $\omega_4$ of the extra dimensions is bounded from a value below $-1$ for a positive $\epsilon$ due to the extra-dimension contribution. More explicitly, The second condition in (\ref{NEC}) can be written as
\begin{align}
\omega_4 \geq -1  - 12 X^2 /\epsilon\,.
\end{align} 
For the large extra-dimension limit ($X\to 0$), the above condition becomes the usual one whose bound is given by a cosmological constant. On the other hand, for a finite extra dimension, the bound of the equation of state $\omega_4$ is less than $-1$. Thus one can say that the extra-dimension volume relaxes the bound of the NEC.

In \cite{KKYH001}, it is explained how an expanding universe together with a dynamical compactification is possible by considering the Maldacena--Nunez no--go theorem in the context of our model~\cite{Maldacena:2000mw}. Even though the theorem can not be applied to an accelerating universe, a slight generalization of the theorem is useful to understand our case. The key quantity of the theorem is the scalar curvature of the (3+1)--dimensional universe. In this perfect-fluid case, the curvature is
\begin{align}\label{Curvature4}
R^{(g)} =6 \left(2H^2+\dot{H}^2\right) = 6\left[ - 4 H \dot{f} -H^2 -X^2 + \epsilon\left( \frac{\omega _3}{2}-\frac{2 \omega _4}{3}+\frac{1}{6} \right) \right]\,.
\end{align}
In the absence of the matter ($\epsilon=0$), a negative value of $H\dot{f}$ is crucial to obtain a positive curvature $R^{(g)}$ that describes an accelerating or expanding universe as in \cite{KKYH001}. So $\dot{f}$ is always negative for positive $H$ or positive for negative $H$. To have an expanding three--space in the universe, $\dot{f}$ must be negative. Thus dynamical compactification is necessary for an accelerating universe. In addition, one may also notice that $H=\dot{f}=0$ case corresponds to $R^{(g)}=0$ with $X=0$. This is nothing but a Minkowski spacetime with an infinite size of extra dimensions. This solution doesn't flow under time evolution and divides the full solution space into positive $\dot{f}$ (or negative $H$) solution set and negative $\dot{f}$ (or positive $H$) solution set.  

On the other hand, the existence of matter relaxes the negativity of $H\dot{f}$ to produce an expanding three--space with positive curvature $R^{(g)}$. Therefore, more diverse cases can happen. The following sections discuss various effects of matter on this 8--dimensional spacetime. We also comment on selected examples in terms of the curvature expression (\ref{Curvature4}).

\section{Instanton Universe with Cosmological Constant}\label{sec3}

In this section, we study the $\omega_3=\omega_4=-1$ case to see how the structure of the solution space changes by a matter. The energy-momentum conservation in (\ref{dynamical00}) for this case indicates that $\epsilon$ is a constant, which is nothing but a cosmological constant. Setting $\epsilon=\Lambda$ gives the following expression for the inverse size:
\begin{align}\label{eq:size}
e^{-2f}=X = 1 - \sqrt{1+2 \dot{f}^2+H^2+4 \dot{f} H-\frac{\Lambda }{3}}\,.
\end{align}
Consequently, the corresponding flow equation becomes
\begin{align}\label{flowLambda}
\left(\ddot{f},\dot{H}\right)=  \left(5 \dot{f} H+2 H^2-\frac{\Lambda }{3}+X,-8 \dot{f} H-2 \dot{f}^2-4 H^2+\frac{2 \Lambda }{3}-2 X\right)\,.
\end{align}
Since the right-hand side contains only $\dot{f}$ and $H$, the solution flow analysis introduced in \cite{Kim:2018mfv} is again convenient in this case. As one can see, the above equation is a first-order differential equation for the vector field $(\dot{f}, H)$. Therefore, the vector field of the right-hand side indicates the direction of change of $(\dot{f}, H)$ for a small time interval $dt$ between $t$ and $t+dt$. In this sense, we regard the right-hand side of (\ref{flowLambda}) as the solution flow. In what follows, we will analyze how this solution flow changes according to the values of the cosmological constant. The curvature of the universe (\ref{Curvature4}) is given by
\begin{align}\label{Rg}
R^{(g)}= 6 \left(\frac{\Lambda}{3}-4 \dot{f} H-H^2-X^2\right)\,.
\end{align}
The quantity $\Lambda/3 - 4 \dot{f}H$ should at least be positive to have positive curvature. According to the value of $\Lambda$, the solution-flow structure shows various dynamics. For instance, even for a negative $\Lambda$, an expanding universe solution is still possible for a negative $\dot{f}$ with $|\dot{f}|$ large enough. This dynamical flow can exist due to the instanton in the extra dimensions. Even though we do not show explicit flows of this case, the qualitative behavior is similar to the behavior of the case without matter \cite{Kim:2018mfv, KKYH001}. Thus, in the following subsection, we focus on the positive $\Lambda$ case.

\subsection{Fixed Points and Various Cosmological Evolution}

To find the behavior of solution flow, or trajectory, using (\ref{flowLambda}), one can utilize the linear stability theory. In the linear stability theory, a given dynamical system with fixed points can be linearized around its fixed points by the so-called Jacobian matrix~\cite{Boehmer:2014vea}. The eigenvalues of the Jacobian matrix, evaluated at the fixed point, reveal information about stability. If all eigenvalues have negative (positive) real values, the fixed point is regarded as stable (unstable). The trajectories around stable (unstable) points are attracted to (repelled by) such fixed points. If at least one of the eigenvalues has a positive real value, the corresponding fixed point would not be stable hence it is a saddle point, which attracts trajectories in some directions, but repels them along others. The fixed point is regarded as unstable if all eigenvalues have positive real values; such points repel all trajectories around it. In addition, if all eigenvalues are complex-valued, the sign of the real part contains information about stability. If the real parts of all eigenvalues are negative (positive), the fixed point is a stable (unstable) spiral. Thus, in this subsection, we first find the fixed points of the flow equation, and then check the asymptotic stability at those fixed points via linear stability analysis.  

In Table~\ref{table:1}, we present all possible fixed points of a system determined by (\ref{flowLambda}) and provide their existence conditions, eigenvalues of the Jacobian matrix, and stability information. 

\begin{table}[h!]
\caption{ The fixed points of (\ref{flowLambda}) and their stability analysis based on the eigenvalues:  Here $A_{\pm}\equiv\sqrt{2\Lambda/3- \left(9\pm\sqrt{81-24 \Lambda }\right)/6}$ and $B_\pm\equiv\sqrt{15 \mp\sqrt{81-24 \Lambda }-4 \Lambda}\,.$ Special numbers appear as $5643/1681\sim 3.357$ and $27/8=3.375$.}
\vspace{3mm}
\begin{adjustbox}{width=\textwidth}
\begin{tabular}{c | c c | c | c | c }
\hline\hline  
Pts & $\dot{f}$ & $H$ & Existence & Eigenvalues &Stability and Inverse size $X$ \\ 
\hline 
& & & & & \\
$A_1^\pm$ & $\pm \sqrt{\frac{\Lambda}{21}}$    & $\pm \sqrt{\frac{\Lambda}{21}}$    & $\Lambda\geq0$ & $\left\{\mp \sqrt{\frac{7}{3}\Lambda},\mp \sqrt{\frac{4}{21}\Lambda}\right\}$ & $A_1^+$ is stable and $A_1^-$ is unstable, \\
& & & & & and $X=0$ for $\Lambda\geq 0$.\\ 
\hline
& & & & & \\
$A_2^\pm$ & $0$  & $\pm \frac{1}{2} \sqrt{\frac{2}{3}\Lambda-\frac{1}{6} \left(9-\sqrt{81-24 \Lambda }\right)}$  & $0<\Lambda\leq {27\over8}$ & $\left\{\pm\frac{3}{4} A_{-} \left(1-\frac{\sqrt{41 B_{-}^2-16 \sqrt{6} B_{-}}}{3 B_{-}}\right), \pm\frac{3}{4} A_{-} \left(1+\frac{\sqrt{41 B_{-}^2-16 \sqrt{6} B_{-}}}{3 B_{-}}\right)\right\}$ & $A_2^\pm$ are saddle for $0<\Lambda\leq {27\over8}$. \\
& & & & &  $0<X\leq\frac{3}{4}$. \\ 
& & & & & \\
\hline
& & & & & \\
 & & &  & & $A_3^+$ is stable and $A_3^-$ is unstable \\
$A_3^\pm$& $0$ & $\pm \frac{1}{2} \sqrt{\frac{2}{3}\Lambda-\frac{1}{6} \left(9+\sqrt{81-24 \Lambda }\right)}$ & $3<\Lambda\leq{27\over8}$ & $\left\{\mp\frac{3}{4} A_+ \left(1-\frac{\sqrt{41 B_+^2-16 \sqrt{6} B_+}}{3 B_+}\right), \mp\frac{3}{4} A_+ \left(1+\frac{\sqrt{41 B_+^2-16 \sqrt{6} B_+}}{3 B_+}\right)\right\}$ &  for $\frac{5643}{1681}<\Lambda < {27\over8}$; \\
& & & & & $A_3^+$ is spirally stable and $A_3^-$ is\\
& & & & & spirally unstable for $3<\Lambda <\frac{5643}{1681}$. \\
& & & & & $\frac{3}{4}\leq X<1$ for $3<\Lambda\leq\frac{27}{8}$. \\
\hline
& & & & & \\
$A_4^\pm$ & $\pm \sqrt{\frac{\Lambda-3}{21}}$ & $\pm \sqrt{\frac{\Lambda-3}{21}}$ & $\Lambda\geq3$ & Indeterminate & $X=1$  \\
& & & & & \\
\hline\hline
\end{tabular}
\end{adjustbox}
\label{table:1} 
\end{table}

The first notable pair of fixed points are the $A_1^\pm$ points. As the table shows, these points exist only for $\Lambda\geq0$, which corresponds to the positive cosmological constant. However, the $\Lambda=0$ value, the vanishing cosmological constant, is excluded because the linear stability theory fails to determine stability when at least one of all eigenvalues of the Jacobian matrix becomes zero \cite{Boehmer:2014vea} \footnote{These vanishing eigenvalues denote $\ddot{f}=\dot{H}=0$ with the fixed points $\dot{f}=H=0$, and the corresponding inverse size is given by $X=0$. Therefore, this is the Minkowski limit with large extra dimensions.}. For $\Lambda>0$, the system is stable at the $A_1^+$ fixed point and unstable at the $A_1^-$ fixed point due to the negative and positive sign of eigenvalues, respectively. 

Next,  the $A_2^\pm$ and $A_3^\pm$ fixed points look similar, but their phenomenological implications are quite different. As for the $A_2^\pm$ fixed points, the existence condition implies that these fixed points can have real values for $0<\Lambda\leq 3.375$, where the upper bound $\Lambda=3.375$ is determined by the real values of $H$ expression in the fixed point column. Within this range of $\Lambda$ values, one obtains two eigenvalues have opposite signs; one of the eigenvalues is positive while the other one is negative. Thus, the $A_2^\pm$ fixed points are always saddle points. On the other hand, the $A_3^\pm$ fixed points exist when the $\Lambda$ takes values within the $3<\Lambda\leq3.375$ range. As is seen in the table, the existence condition for $A_3^\pm$ points is tighter than that of the $A_2^\pm$ points. The system is stable at the $A_3^+$ point and unstable at the $A_3^-$ point when $3.357\lesssim\Lambda\leq 3.375$. In the $3<\Lambda< 3.357$ range, the eigenvalues are complex-valued hence the sign of their real parts reveals that the $A_3^-$ point is spirally unstable and the $A_3^+$ point is spirally stable.
 \begin{figure}[H]
    \centering
    \includegraphics[width=\textwidth]{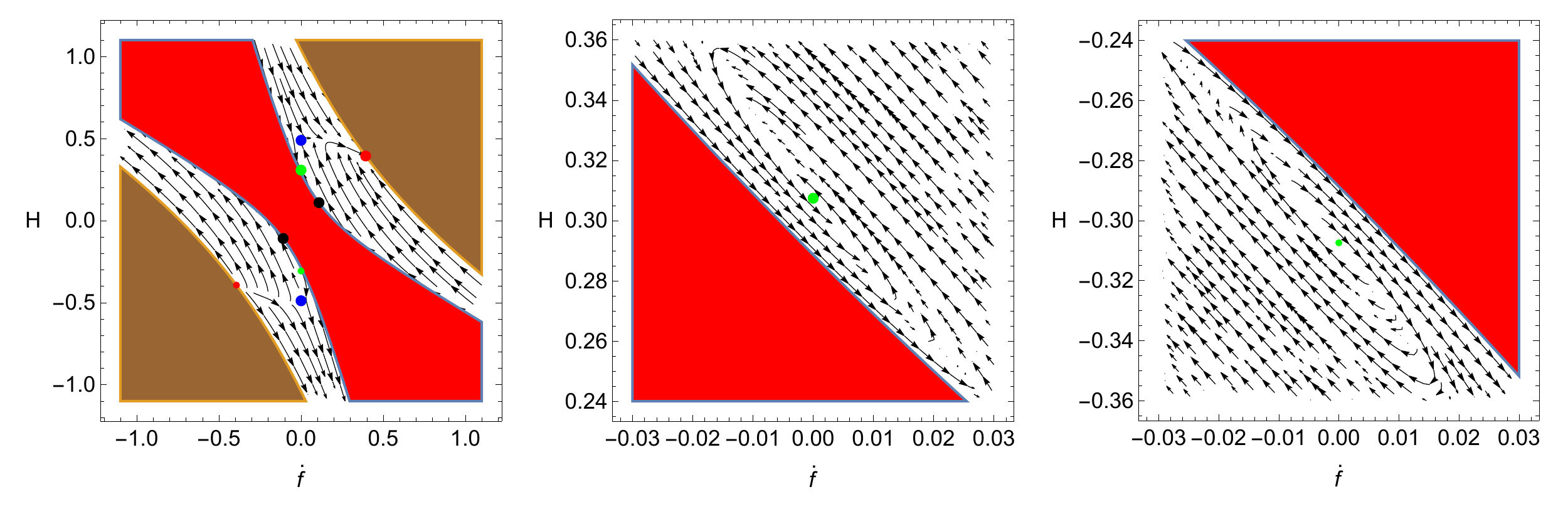}
    \caption{Phase space diagrams of the autonomous system on a $\dot{f}$ vs. $H$ plane described by (\ref{flowLambda}). We take $\Lambda=13/4$. The shaded regions indicate the parameter space excluded by the positivity of $X$ (\emph{brown}) and the quantity under the square root in $X$ in (\ref{eq:size}) to be real-valued (\emph{red}). The red, blue, green, and black dots represent the $A_1^\pm$, $A_2^\pm$, $A_3^\pm$, and $A_4^\pm$ fixed points in Table~\ref{table:1}, respectively.} \label{fig:01}
\end{figure}

In Figure~\ref{fig:01}, we plot the phase portraits of our dynamical system described by (\ref{flowLambda}) in a $\dot{f}$ vs. $H$ plane. To present all fixed points listed in Table~\ref{table:1} in one plane, we set $\Lambda=3.25$. The brown- and red-shared regions indicate the parameter space excluded by the $X$ to be positive and real-valued, respectively. The red, blue, green, and black dots represent the $A_1^\pm$, $A_2^\pm$, $A_3^\pm$, and $A_4^\pm$ fixed points, respectively.  As we discussed earlier, the $A_1^-$ fixed point, or the small red dot, is the unstable fixed point hence trajectories around it are repelled by this point, see the outgoing arrows in the figure. The $A_1^+$ fixed point, the large red dot, conversely, is the stable fixed point, which attracts trajectories around it as indicated by ingoing arrows. Table~\ref{table:1} also shows that the $A_2^\pm$ fixed points are always the saddle points. The blue dots in the figure represents the $A_2^\pm$ points, and, as saddle points, they attract trajectories in some directions, but repel them along others. The chosen value $\Lambda=3.25$ falls in the range $3<\Lambda\lesssim 3.357$. Thus, the $A_3^-$ fixed point, the small green dot, is an unstable spiral, while the $A_3^+$ fixed point, the large-green dot, is a stable spiral. For the unstable spiral, the arrows of trajectories toward outward, while the arrows of trajectories toward inward for the stable spiral, see middle and right panels in Figure~\ref{fig:01}, respectively. 

The time-dependent Hubble parameter $H(t)$ and the cosmological constant $\Lambda$, in this case, are independent due to the choice of the initial condition, while those are directly related to each other in a de Sitter spacetime without extra dimensions. Thus these two quantities are useful observables to show the role of extra dimensions. To see how they are related to physical observables, we need to express the physical Hubble parameter and the physical cosmological constant in terms of $\Lambda$ and $H(t)$ using (\ref{reScale00}). If we use the length scales given by $16\pi G_{(8)}= 2l_p^6$ and $g_{\text{YM}}^2=l_{\text{YM}}^4$, the physical cosmological constant $\Lambda_p$ of 8--dimensional spacetime can be written as
\begin{align}
\Lambda_p l_p^2 =  2\Lambda  \left( \frac{l_{\text{YM}}}{l_p} \right)^4\,.
\end{align}
Also, the physical Hubble parameter function in the unit of the Planck scale can be written by
\begin{align}
H^2_p(t) l_p^2 =  H^2(t)  \left( \frac{l_{\text{YM}}}{l_p} \right)^4\,.
\end{align}\label{Hp}
Thus the ratio of the Hubble parameter to the cosmological constant is determined by
\begin{align}
\frac{H_p^2}{\Lambda_p} = \frac{H^2}{2\Lambda}\,.
\end{align}
In a usual de Sitter spacetime without extra dimension, this ratio is just a constant. As opposed to this pure de Sitter case, the ratio is determined by the dynamical field $H(t)$ based on the flow equation (\ref{flowLambda}). We will discuss this quantity in a specific example, which is a de Sitter fixed point, later. The following two subsections are devoted to some intriguing cosmological evolutions. One is the bouncing universe solution, which is ubiquitous in the solution flow, and the other is the (spirally) stable de Sitter solution  ($A_3^+$) appearing in Table \ref{table:1}.

\subsection{Bouncing universe solution}

Our model is based on the interplay between the universe and extra dimensions through the Einstein-matter theory. The volume-preserving nature of Einstein gravity is expected to give rise to seesaw-like dynamics of the universe and the extra dimensions as a very common time evolution. In the universe view or the extra dimension view, these dynamics can be regarded as bouncing universes or bouncing extra dimensions. This subsection presents bouncing solutions and conditions for the bouncing behavior.  

\begin{figure}[H] 
    \centering
    \includegraphics[width=\textwidth]{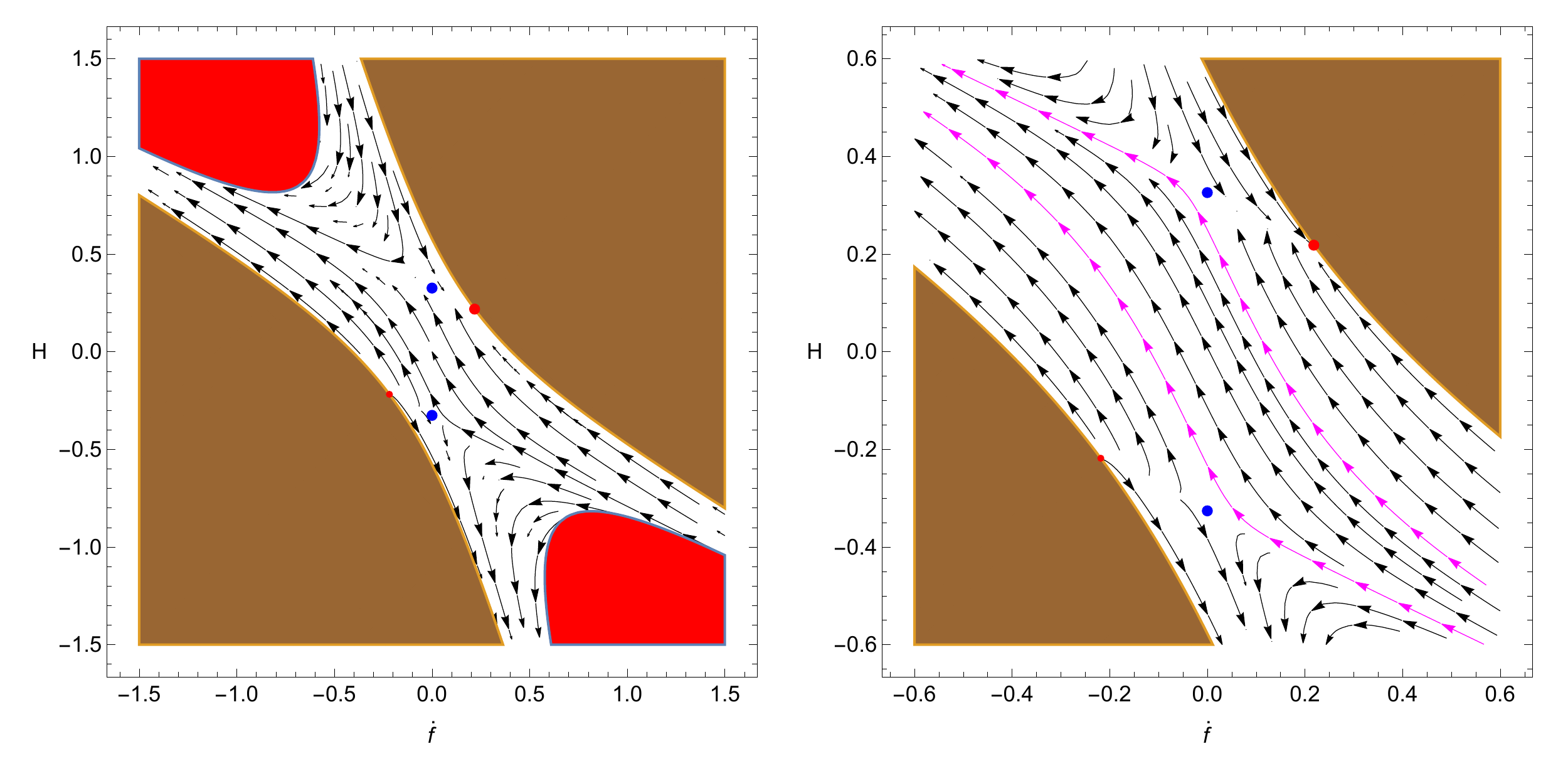}\\
    \includegraphics[width=7cm]{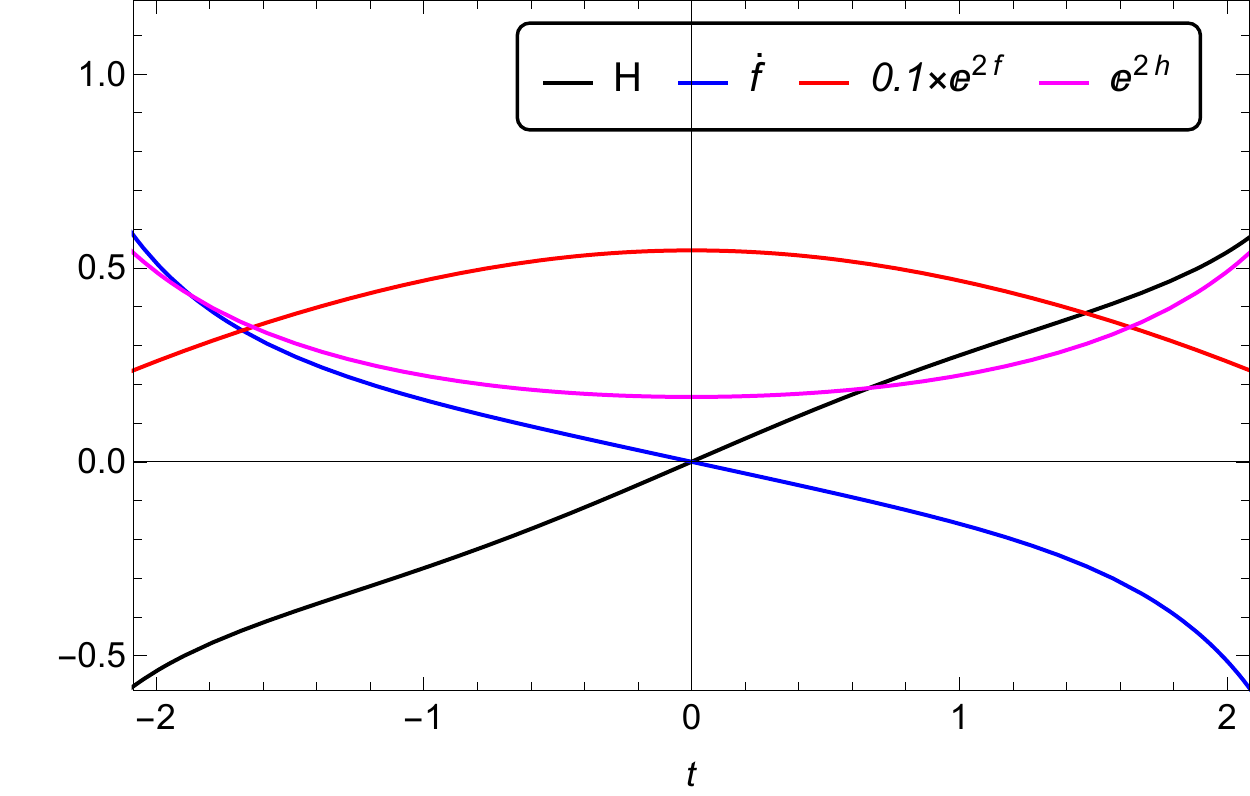}~
    \caption{\emph{Top}: Phase space diagrams on a $\dot{f}$ vs. $H$ plane  of a system described by (\ref{flowLambda}) for $\Lambda=1$. The shaded regions have the same meaning as Figure~\ref{fig:01}. The red and blue dots represent the $A_1^\pm$ and $A_2^\pm$ fixed points in Table~\ref{table:1}, respectively. The magenta lines indicate particular bouncing solutions. \emph{Bottom}: The time evolution of $H$, $\dot{f}$, $e^{2f}$ and $e^{2h}$ for the same $\Lambda=1$ showing the particular bouncing solution at $t=t_b(=0)$, \emph{i.e.,} $\{H, \dot{f}\}=0$.} \label{fig:02}
\end{figure}

To describe the bouncing stage, we plot the $H$ vs. $\dot{f}$ phase diagram in the upper panel and the time evolution of $H(t)$ and $\dot{f}(t)$, as well as the sizes of the three space of the universe -- $e^{2h(t)}$ and the extra dimensions -- $e^{2f(t)}$, in the bottom panel of Figure~\ref{fig:02}, respectively, for $\Lambda=1$. The shaded regions are the same as Figure~\ref{fig:01}. Also, the red and blue dots represent the $A_1^\pm$ and $A_2^\pm$ fixed points in Table~\ref{table:1}, respectively. The magenta lines indicate particular bouncing solutions, where a trajectory with $H<0$, initially, evolves to $H>0$ by crossing zero at $t=t_b$. The bouncing time $t_b$ is different for different solutions. At the same time as the $H$ value increases from a negative to a positive value,  the $\dot{f}$ value decreases from a positive to a negative value. Thus, for the bounce to occur, there must at least be a trajectory that crosses the origin, $\{\dot{f}, H\}=\{0,0\}$, as is shown in the bottom panel of Figure~\ref{fig:02}. In addition to the time evolution of $H$ and $\dot{f}$, we also present the time evolution of $e^{2f}$ and $e^{2h}$. The figure shows the size of the extra dimension (red) grows in time and reaches the maximum size at $t=t_b$. After the bouncing, the size of the extra dimension shrinks, whereas the size of the three spaces (magenta) increases as expected.  

The growing $H$ parameter also indicates the $\dot{H}>0$ during the bounce stage. More explicitly, the solution flow at the origin ($\dot{f}=H=0$) is
\begin{align}
\left\{\ddot{f},\dot{H}\right\}_{\dot{f}=H=0}=\left\{1 -\frac{\Lambda}{3}-\sqrt{1-\frac{\Lambda}{3}}, -2 \left(1 -\frac{\Lambda}{3}-\sqrt{1-\frac{\Lambda}{3}}\right)\right\}\,,
\end{align}
and the slope at the origin is given by $dH/d\dot{f}=-2$ with $X=1-\sqrt{1-\Lambda/3}$. Thus, for $0<\Lambda<3$, there exists always bouncing flows.

When $\Lambda=3$, no bounce stage occurs because the two separate red regions in the figure intersect only at the origin $\{\dot{f}, H\}=\{0,0\}$. A solution with $\Lambda=3$ results in vanishing  $\dot{H}$ and $H$ as well as $\ddot{f}$ and $\dot{f}$, so this solution does not evolve. Therefore, this solution is isolated and describes $\mathbb{R}^{1,3}\times\mathbb{S}^4$. The stability of such a solution was studied in \cite{Randjbar-Daemi:1983xth}~\footnote{See the curvature expression (\ref{Rg}). This geometry has $X=1$. As we discussed earlier, a quantum gravity effect must be considered. Our argument and this study are achieved without any assumption on $g_{\text{YM}}^2$.}.

In most cases of bouncing models in cosmology, it is hard to obtain non-singular time evolution without violating the NEC~\cite{Novello:2008ra, Battefeld:2014uga, Lilley:2015ksa, Brandenberger:2016vhg}. However, this bouncing dynamics fulfills the NEC and gives rise to regular time evolutions. A difference from other bouncing models is that our model has flexible extra dimensions which relax the NEC (\ref{NEC}).

\subsection{de Sitter Universe}

Let us discuss another interesting solution given by stable fixed points. When $\Lambda>0$, the stable fixed points are the $A_1^+$ and $A_3^+$ in Table~\ref{table:1}. However, at the $A_1^+$ point, the size of an extra dimension is infinite, i.e., $X=0$; therefore, the total 8--dimensional spacetime undergoes a decompactification process approaching the $A_1^+$ point. Thus, we exclude the $A_1^+$ point from our interest and focus more on the $A_3^+$ point instead. 

\begin{figure}[h!]
    \centering
    \includegraphics[width=\textwidth]{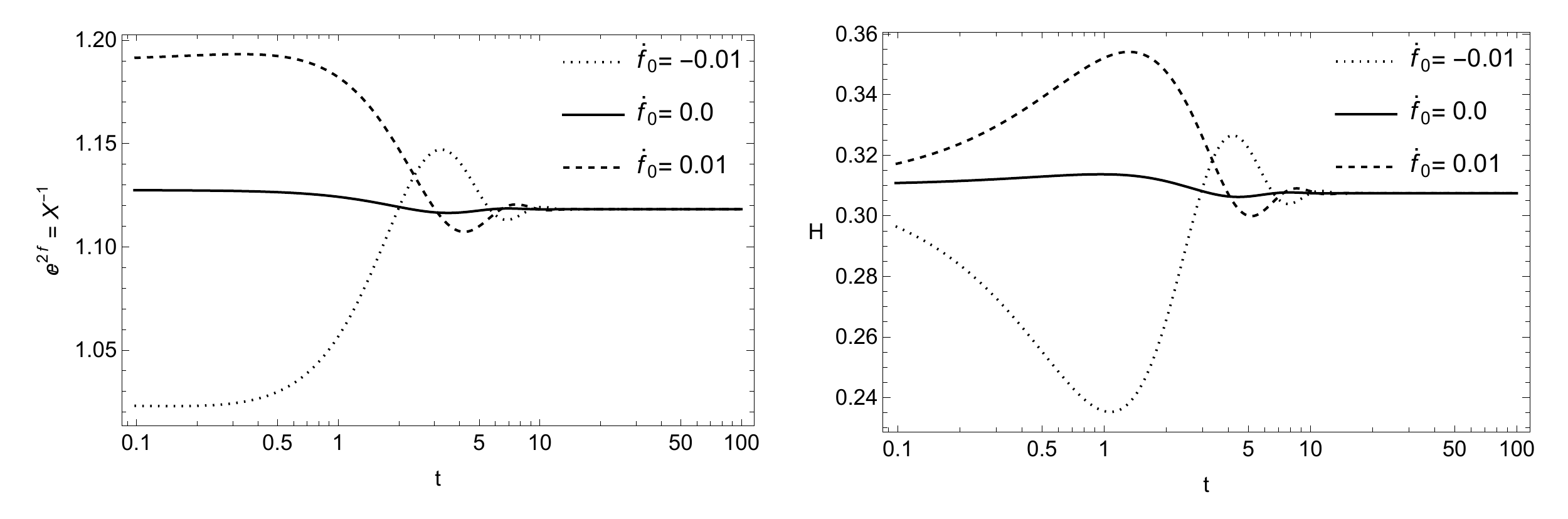}\\
    \includegraphics[width=\textwidth]{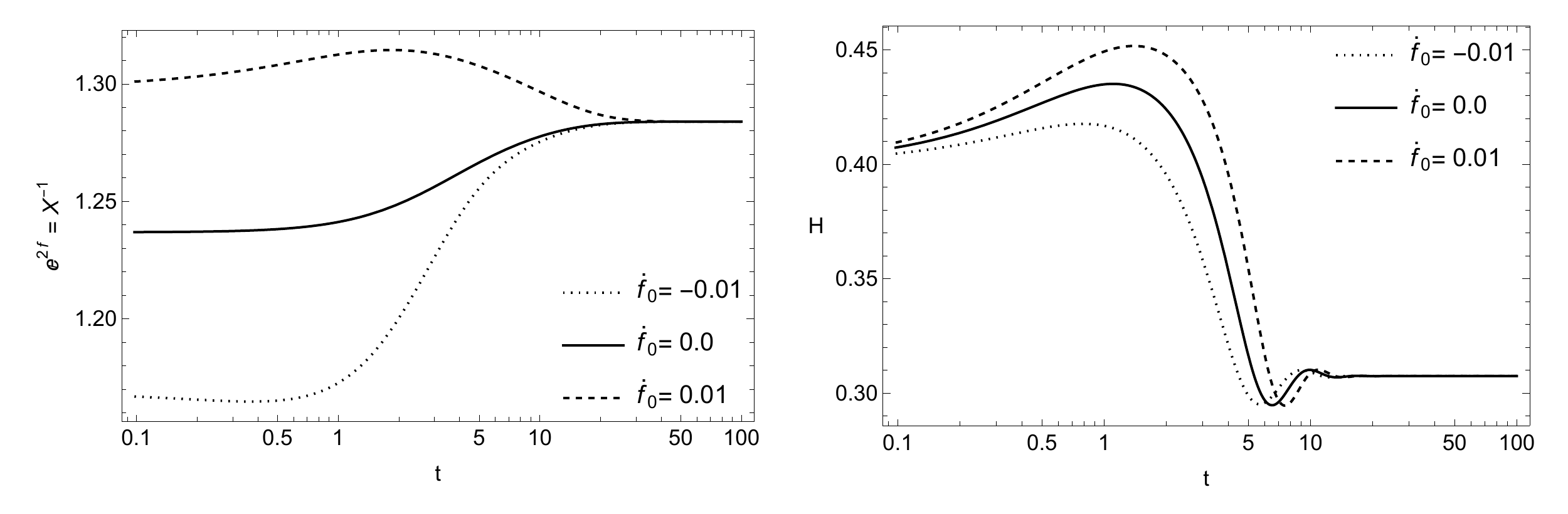}
    \caption{Dynamical evolution of the size of extra dimensions in (\ref{eq:size}) and the growth rate of the universe; left and right columns, respectively. When $\Lambda=3.25$ (upper panels), the universe is described by a spirally stable point, and when  $\Lambda=3.37$ (lower panels), the universe is described by a stable point. }
    \label{fig:03}
\end{figure}
 
At the $A_3^+$ fixed point, the size of the extra dimensions is stabilized, and the universe becomes a de Sitter spacetime~\footnote{Since there is a positive cosmological constant $\Lambda$, this situation can evade the Maldacena--Nunez no--go theorem \cite{Maldacena:2000mw}.}. The corresponding $X$ becomes 
\begin{align}
X=1-\frac{1}{2} \sqrt{-\frac{2 \Lambda }{3}-\sqrt{\frac{9}{4}- \frac{2}{3} \Lambda}+\frac{5}{2}}\,,
\end{align}
at this fixed point. Since this inverse size is close to 1 ($1>X>3/4$), the quantum gravity effect should correct this value. Leaving such a concern as future work, we continue our discussion on this scale. The 4--dimensional Newton constant and the corresponding Planck scale are given by
\begin{align}
16\pi G_N = 6 \times \frac{16\pi G_{(8)}}{\pi^4 \zeta_4^4}e^{-4 f}~,~l_{p,4}^2 =\frac{3}{8\pi^2} X^2 \left(\frac{l_{\text{YM}}}{l_p}\right)^8 l_p^2\,,
\end{align}  
where we used the scaling (\ref{reScale00}). Employing this quantity, the Hubble parameter becomes
\begin{align}\label{Hubble00}
H_p^2 l_{p,4}^2 = \frac{3}{8 \pi^2}\left(\frac{l_{\text{YM}}}{l_p}\right)^{12} H^2 X^2 \,,
\end{align}
where we used (\ref{Hp}). In the unit of the 4--dimensional Planck scale, this Hubble parameter depends on the ratio of the coupling-length scales and $H^2 X^2$. Plugging the fixed point $A_3^+$ into $H$ and $X$, we obtain the following expression:
\begin{align}
\left.H^2 X^2\right|_{A_3^+} = \frac{\left(12-\sqrt{6} \sqrt{-4 \Lambda -\sqrt{81-24 \Lambda }+15}\right)^2 \left(4 \Lambda -\sqrt{81-24 \Lambda }-9\right)}{3456}\,.
\end{align}
This is a monotonically increasing function of $\Lambda$, which ranges from $0$ to $27/256\sim 0.105$. When the spacetime approaches $\Lambda=3$ and $X=1$, this function is closed to zero. 

Since our approach is a toy model, we comment that we do not know any fundamental theory which can be related to our model. If, however, we naively consider a relation $G_{(8)}=g_{\text{YM}}^2  l_s^2$ between the scales appearing in (\ref{Hubble00}), where $l_s$ is an unknown fundamental scale, then the Hubble parameter can be estimated as follows:
\begin{align}\label{H_p l_p4}
H_p^2 l_{p,4}^2 =\frac{24}{(8\pi)^5}\left( \frac{l_p}{l_s} \right)^6 H^2X^2 \simeq 2.39 \times 10^{-7} \left( \frac{l_p}{l_s} \right)^6 \,,
\end{align} 
where we set $H^2X^2=0.1$ close to the maximum value 27/256 of $H^2 X^2$. If we also assume that $l_p$ and $l_s$ are in the same or similar order, the Hubble parameter gets the small value, in terms of the 4--dimensional Planck length, because of the smallness of the numerical factors and the smallness of $H^2 X^2$. Since the inverse size of the extra dimensions $X$ is close to 1, it would be interesting to see how this quantity $H^2_p l_{p,4}^2$ changes by a quantum gravity effect. We leave this as a future study. This estimation can explain the toy version of the cosmological constant problem in our model. It is notable that $H_p^2 l_{p,4}^2$ is extremely small while the 8--dimensional cosmological constant $\Lambda_p l_p^2=\Lambda(l_p/l_s)^2/(4\pi)\sim 0.25 (l_p/l_s)^2$ is the order of one. 

Now let us find the dynamics before arriving at the de Sitter fixed point. We numerically integrate (\ref{flowLambda}) to plot the dynamical evolution of the $H(t)$ and the size of an extra dimension $e^{2f}=X^{-1}$ in Figure~\ref{fig:03}. Two different values of $\Lambda$, namely $\Lambda=3.25$ (upper) and $\Lambda=3.37$ (lower), are considered for plotting the figure, where we show different dynamics towards the different stable fixed points.  The full geometry settles at the $A_3^+$ fixed point, which can be either the stable or stable spiral fixed point according to the $\Lambda$ value. The $A_3^+$ point is a spirally stable fixed point if $3<\Lambda\lesssim 3.357$ and is a stable fixed point if $3.357\lesssim \Lambda \leq 3.375$, see Table~\ref{table:1}. Thus, when $\Lambda=3.25$, both the $H(t)$ and $X^{-1}(t)$ oscillate several times before settling at the constant values. In contrast, there is less oscillation that occurs when $\Lambda=3.37$. The value of both $\dot{f}$ and $H$ is constant at $A_3^+$ fixed point, and the size of the extra dimension is finite after all. The oscillation dynamics can be considerable since it can generate another matter production, such as a reheating process. Even though it is interesting, we plan to study it after understanding a quantum gravity correction.

\section{Instanton Universe with More General Matter}

The previous section discussed various effects of the cosmological constant on cosmological evolution. Thus, we discuss the effects of more general matter within our model, rather than the cosmological constant, in this section. We first consider matter energy that is given in terms of the size of extra dimensions, and then a scalar field with potential next. 

\subsection{Matter with volume-dependent energy density}
As a first try for a nontrivial matter, we consider a matter satisfying the NEC bound in (\ref{NEC}). Thus, in this subsection, we take $\omega_3 =-1$ and $\omega_4 = -1 - c_0 X^2/\epsilon$, where $c_0\leq12$ is an arbitrary constant, into consideration. After substituting these expressions into the energy-momentum conservation law, we solve (\ref{dynamical00}) to obtain
\begin{align}
\epsilon = \epsilon_0 - c_0 X^2\,,\label{energydensity}
\end{align}
where $\epsilon_0>0$ is the constant energy density when the size of the extra dimensions is infinite, {\it i.e.}, $X=0$. To guarantee the positive energy density, one has to impose $\epsilon_0\geq c_0 X^2$ by considering the validity of classical solutions. However, negative energy also satisfies the NEC. After plugging this energy density into (\ref{flowEq0}), the equation of motion becomes
\begin{align}
&H^2+4 \dot{f} H+2 \dot{f}^2+\frac{c_0-3}{3} X^2+2 X-\frac{\epsilon _0}{3}=0\,,\label{eq:consteq}\\
&4 \ddot{f}+8 \dot{f} H+10 \dot{f}^2+3 H^2+2 \dot{H}+(c_0-3) X^2+6 X-\epsilon _0=0\,,\\
&3 \ddot{f}+9 \dot{f} H+6 \dot{f}^2+6 H^2+3 \dot{H}+3 X-\epsilon _0=0\,,
\end{align} 
where the first equation is again a constraint equation. The constraint equation has only one root for $X$ when $c_0=3$ and two roots when $c_0\neq 3$. 

Let us consider the $c_0\neq3$ case first, for which we find two roots, but the root we are interested in is
\begin{align}\label{eq:forX}
e^{-2f}=X=\frac{3}{c_0-3} \left[\sqrt{\frac{c_0-3}{9}\left(-12 \dot{f} H-6 \dot{f}^2-3 H^2+\epsilon _0\right)+1}-1\right]\,.
\end{align}
and the other root is excluded because it gives a negative value. From (\ref{eq:forX}), the physical conditions read as $1>X\geq 0$ and $(c_0-3)(-12 \dot{f} H-6 \dot{f}^2-3 H^2+\epsilon _0)/9+1\geq 0$, which can be reduced to
\begin{align}
c_0+3>\epsilon _0-12 \dot{f} H-6 \dot{f}^2-3 H^2 \geq 0\,.\label{conforX}
\end{align}
The first inequality corresponds to $X<1$.  Using this solution, the flow equation is given by
\begin{align}\label{flowNEC}
\left\{\ddot{f},\dot{H} \right\}=\left\{5 \dot{f} H+2 H^2+X-\frac{\epsilon _0}{3} ,-8 \dot{f} H-2\dot{f}^2-4 H^2-2 X+\frac{2 \epsilon _0}{3} \right\}\,.
\end{align}
We provide the fixed points of the flow equation (\ref{flowNEC}) in Table~\ref{table:2} and present them in the phase space diagram in the upper left panel of Figure~\ref{fig:03}, where $\epsilon_0=1$ and $c_0=12$. 
\begin{table}[h!]
\caption{ The fixed points of (\ref{flowNEC}) and their stability analysis based on the eigenvalues:  where $\alpha_{\pm}\equiv \sqrt{\frac{\epsilon _0}{6}\pm\frac{1}{8(c_0-3)}\left(\sqrt{81+ 8 \epsilon _0(c_0-3)}\pm9\right)}$ and $\beta_{\pm}=\sqrt{12(c_0-3) \alpha_{\pm}^2+9}$. The existence indicates classically reliable parameter regions. }
\vspace{3mm}
\begin{adjustbox}{width=\textwidth}
\begin{tabular}{c | c c | c | c | c }
\hline\hline  
Pts & $\dot{f}$ & $H$ & Existence & Eigenvalues &Stability\\ 
\hline 
& & & & & \\
$C_1^{-,\pm}$ & 0    & $-\alpha_{\pm}$    & $0<\epsilon_0\leq \frac{81}{8(3-c_0)}$ for $c_0 <3$;  & $\frac{\alpha_{\pm}}{2}\times\left\{ 3-\sqrt{\frac{41\beta_{\mp}-48}{\beta_\mp}},  3+\sqrt{\frac{41\beta_{\mp}-48}{\beta_\mp}}  \right\}$ & $C_1^{-,\pm}$ are saddle.\\
& & & $\epsilon_0>0$ for $c_0 >3$  & &  \\
& & & & & \\
\hline
& & & & & \\
$C_1^{+,\pm}$ & 0    & $\alpha_{\pm}$    & $0<\epsilon_0\leq \frac{81}{8(3-c_0)}$ for $c_0 <3$; & $-\frac{\alpha_{\pm}}{2}\times\left\{ 3+\sqrt{\frac{41\beta_{\mp}-48}{\beta_\mp}},  3-\sqrt{\frac{41\beta_{\mp}-48}{\beta_\mp}}  \right\}$ & $C_1^{+,\pm}$ are saddle.\\
& & & $\epsilon_0>0$ for $c_0 >3$ & &   \\
& & & & & \\
\hline
& & & & & \\
$C_2^{\pm}$& $\pm \sqrt{\frac{\epsilon_0}{21}}$ & $\pm \sqrt{\frac{\epsilon_0}{21}}$ & $ \epsilon_0>0$ & $\{\mp \sqrt{\frac{7\epsilon_0}{3}},\mp \sqrt{\frac{4\epsilon_0}{21}} \}$ & $C_2^{+}$ is stable, while $C_2^{-}$ is unstable.  \\
& & & & &  \\
\hline\hline
\end{tabular}
\end{adjustbox}
\label{table:2}
\end{table}
The table shows that the fixed $C_1^{-,\pm}$ and $C_1^{+,\pm}$ points are saddle points for their two eigenvalues taking the opposite signs. This result is confirmed by the figure where trajectories nearby the black dots, corresponding to $C_1^{-,+}$ and $C_1^{+,-}$, are attracted to them, but eventually pass by them, indicating that these points are not stable attractor points. As for the $C_2^+$ point, the stability column of Table~\ref{table:2} presents that this fixed point is the stable attractor, which is then confirmed in the figure where trajectories around the $C_2^+$  point, the larger red dot, are attracted to it. Conversely, the trajectories around the $C_2^-$ point, the smaller red dot, towards an outward direction, implying that the $C_2^+$ is an unstable repeller. In the figure, the fixed $C_1^{-,-}$ and $C_1^{+,+}$ points are not presented as they locate inside the excluded region (brown) of $X<0$. The red-shaded region is also excluded because, inside the region, we have $X\geq1$. The green line in Figure~\ref{fig:03} indicates that our model makes the bouncing universe scenario possible. The time evolution of quantities, including $H$, $\dot{f}$, $e^{2f}$, and $e^{2h}$, are plotted in the bottom-left panel, where the bounce occurs at $t=0$ and is determined by the flow at $\{\dot{f}, H\}=0$. Thus, the negative (positive) $t$ means before (after) the bounce.  The evolutions of the $e^{2f}$ and $e^{2h}$ are similar to the ones presented in Figure~\ref{fig:01}; decreasing (increasing) extra dimension (three spaces) after the bounce. 

When $\{\dot{f}, H\}=0$, we obtain 
$X= (\sqrt{9+(c_0-3)\epsilon_0}-3)/(c_0-3)$.
Consequently,~(\ref{conforX}) becomes $c_0+3>\epsilon_0\geq0$. Furthermore, the flow equations (\ref{flowNEC}) become
\begin{align}
\left\{\ddot{f},\dot{H}\right\}_{\dot{f}=H=0} =\frac{9-3\sqrt{9+\epsilon_0(c_0-3)}+\epsilon_0(c_0-3)}{(c_0-3)}\times\left\{-\frac{1}{3},\frac{2}{3}\right\}\,,
\end{align}
and the flow slope is given as $dH/d\dot{f}=-2$. 

\begin{figure}[h!] 
    \centering
    \includegraphics[width=\textwidth]{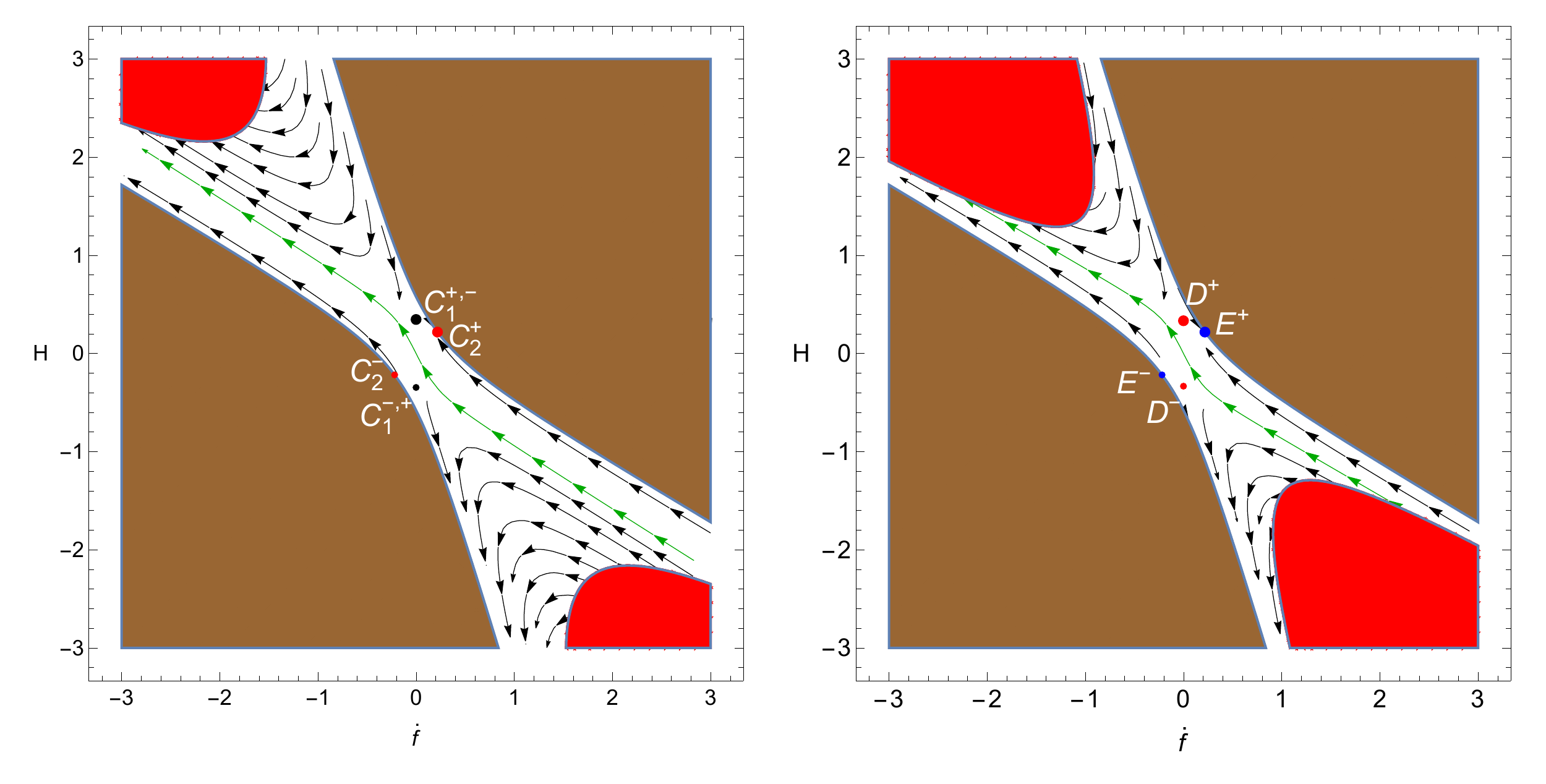}~\\
    \includegraphics[width=\textwidth]{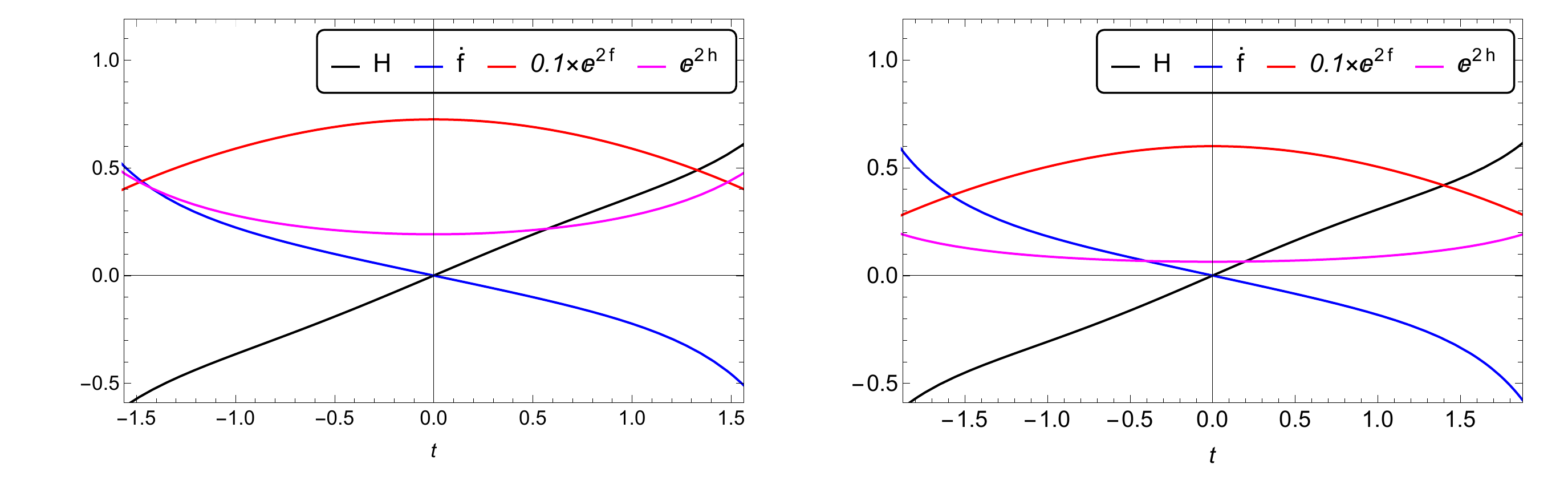}~
    \caption{\emph{Top}: Phase space diagrams on a $\dot{f}$ vs. $H$ plane of a system described by (\ref{flowNEC}) and (\ref{flowNECc3}), left and right, respectively, for $\epsilon_0=1$. The shaded regions have the same meaning as Figure~\ref{fig:01}. The dots on each panel denote the fixed points in Table~\ref{table:2} and Table~\ref{table:3}. The green lines indicate particular bouncing solutions. \emph{Bottom}: The time evolution of $H$, $\dot{f}$, $e^{2f}$ and $e^{2h}$ along the green lines in the respective top panels. The bounce occurs at $t=0$. } \label{fig:04}
\end{figure}

Let us now discuss the $c_0=3$ case. The constraint equation~(\ref{eq:consteq}) has the only root for $X$,
\begin{align}
e^{-2f} =X = -\dot{f}^2 - 2\dot{f} H -\frac{1}{2} H^2+\frac16\epsilon_0\,.
\end{align}
Consequently, the flow equations read
\begin{align}
\{ \ddot{f}, \dot{H} \}= \{ -\dot{f}^2 +3\dot{f} H +\frac32H^2-\frac16\epsilon_0, -4\dot{f} H-3H^2+\frac13\epsilon_0 \}\,.\label{flowNECc3}
\end{align}
Having obtained the dynamical flow equation, we can get the fixed points of the flow equations following the same procedure as the $c_3\neq0$ case. Table~\ref{table:3} presents all the fixed points, their existence, and their stability information, while the upper-right panel of Figure~\ref{fig:03} shows the solution flow diagram for $\epsilon_0=1$.  The figure confirms the statement addressed in Table~\ref{table:3}. When $c_0=3$, we have four fixed points. The red dots, corresponding to the $D^\pm$ fixed points, in the figure are saddle points; trajectories pass through these points, allowing the universe to experience the bounce. The larger blue dots correspond to the $E^{+}$ fixed points, which is the stable attractor, and the trajectories near this fixed point are attracted to it. In contrast, trajectories near $E^-$ fixed point, the smaller blue dot, move away from it, indicating this fixed point is an unstable repeller. The particular bouncing solution is also plotted in the bottom-right panel, where the bounce occurs at $t=0$ and is determined by the flow at $\{\dot{f}, H\}=0$. The slope and the inverse size are $dH/d\dot{f}=-2$ and $X=\epsilon_0/6$, respectively.
\begin{table}[h!]
\caption{ The fixed points of (\ref{flowNECc3}) and their stability analysis based on the eigenvalues. The existence indicates classically reliable parameter regions. }
\vspace{3mm}
\begin{adjustbox}{width=\textwidth}
\begin{tabular}{c | c c | c | c | c }
\hline\hline  
Pts & $\dot{f}$ & $H$ & Existence & Eigenvalues &Stability\\ 
\hline 
& & & & & \\
$D^{\pm}$ & 0    & $\pm \sqrt{\epsilon_0}/3$    &  $\epsilon_0>0$ & $\{\mp \left(\sqrt{33}+3 \right)\sqrt{\epsilon_0}/6, \pm\left(\sqrt{33}-3 \right)\sqrt{\epsilon_0}/6\}$ & $D^{\pm}$ are saddle.\\
& & &  & &  \\
\hline
& & & & & \\
$E^{\pm}$ & $\pm\sqrt{\epsilon_0/21}$   & $\pm\sqrt{\epsilon_0/21}$    & $\epsilon_0>0$ & $\left\{ \mp \sqrt{7\epsilon_0/3}, \mp\sqrt{4\epsilon_0/21}\right\}$ & $E^{+}$ is stable, while $E^{-}$ is unstable.\\
& & & & &  \\
\hline\hline
\end{tabular}
\end{adjustbox}
\label{table:3}
\end{table}

The geometry for the stable fixed point has 4-dimensional de Sitter and expanding extra dimensions. Thus, this stable-de Sitter-fixed point does not have a stable size of the extra dimensions. Unlike the cosmological constant case, where $\dot{f}$ is positive. To understand this, we write down the 4-dimensional curvature as 
\begin{align}
R^{(g)}= -24 \dot{f} H-6 H^2+2 \epsilon_0+2(c_0-3)X^2\,.
\end{align}
One can notice here that there are expansion sources given by $2\epsilon_0+2c_0X^2$, which allows the universe to accelerate even for a positive $\dot{f}H$.  

\subsection{Scalar for de Sitter spacetime}

This section introduces a real scalar field, with only time dependence, as a matter field. This case can be regarded as a homogeneous excitation of the scalar field on the extra dimensions. The energy-momentum tensor is given by
\begin{align}
T_{MN}^\phi = \left(\frac{1}{2}\dot{\phi}^2 + U(\phi) \right)u_M u_N + \left(\frac{1}{2}\dot{\phi}^2 - U(\phi) \right)\left(P_{MN}^{(3)}+ P_{MN}^{(4)} \right)\,.
\end{align}
Thus the energy density and the pressures can be read off as follows:
\begin{align}
\mathcal{E}=\left(\frac{1}{2}\dot{\phi}^2 + U(\phi) \right)~,~\mathcal{P}_{(3)}=\mathcal{P}_{(4)}= \left(\frac{1}{2}\dot{\phi}^2 - U(\phi) \right)\,.
\end{align}
For convenience, one may use the scaled energy density and the equation of state as 
\begin{align}
\epsilon=8\pi G_{(8)}\left(\frac{1}{2}\dot{\phi}^2 + V(\phi) \right)  ~,~\omega_3 = \omega_4 = \frac{\frac{1}{2}\dot{\phi}^2 - V(\phi)}{\frac{1}{2}\dot{\phi}^2 + V(\phi)}\,,
\end{align}
where we define $V(\phi)\equiv \zeta_4^4 U(\phi)/(48\zeta_0^2)$. Then, the conservation law becomes
\begin{align}\label{ScalarEOM}
\frac{d}{dt}\left(\frac{1}{2}\dot{\phi}^2 + V(\phi) \right) =- \left(3H + 4\dot{f}\right)\dot{\phi}^2\,.
\end{align}
On the other hand, the NEC (\ref{NEC}) of this case is given by
\begin{align}
\epsilon=8\pi G_{(8)}\left(\frac{1}{2}\dot{\phi}^2 +V\right) \geq 0\,.
\end{align} 
Thus, the NEC becomes a weak energy condition. The energy density consists of the kinetic part and the potential part. When the potential energy is dominant, {\it i.e.}, $\dot{\phi}\to 0$ limit, such a situation is the same with the cosmological constant matter case in section \ref{sec3}. In the general case, we can use (\ref{dynamical00}) together with $\omega_3=\omega_4$. 

The flow equation (\ref{dynamical00}) with $\omega_3=\omega_4$ determines dynamics of cosmological solutions for given $\omega_3(t)$ and initial conditions $\dot{f}(t_i)$, $H(t_i)$ and $\epsilon(t_i)$, where $t_i$ is a certain initial time. Here, $\omega_3$ and $\epsilon(t_i)$ have the information of the potential $V(\phi)$. We invert the equation of state and the energy density into the kinetic energy and the potential as
\begin{align}\label{K-V}
8\pi G_{(8)}\dot{\phi}^2= \left(1 +\omega_3\right)\epsilon~~,~~16\pi G_{(8)}V=\left(1-\omega_3\right)\epsilon\,,
\end{align}
where one can notice that the NEC with $\omega_3\geq-1$ guarantees the positivity of the kinetic energy. Assuming this condition and a positive energy density, the scalar field can be integrated as follows:
\begin{align}\label{scalar integration}
\phi(t)= \phi(t_i) + \frac{1}{\sqrt{8\pi G_{(8)}}} \int_{t_i}^t dt' \sqrt{(1+\omega_3(t'))\epsilon(t')}\,.
\end{align}
The potential $V(\phi)$ can be obtained from (\ref{K-V}) with the expression  (\ref{scalar integration}).  
\begin{figure}[h!] 
    \centering
    \includegraphics[width=\textwidth]{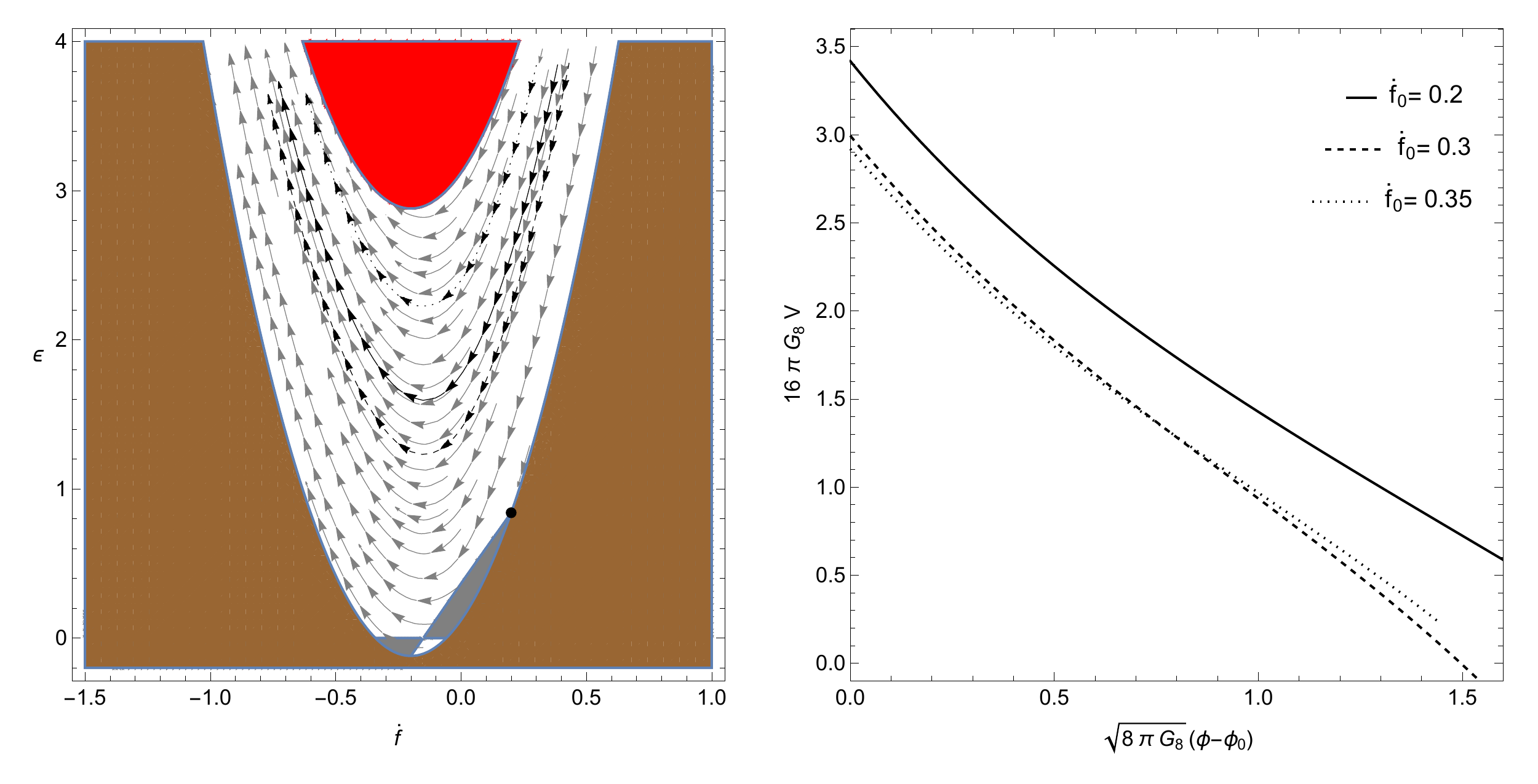}~~~~\\
    \includegraphics[width=\textwidth]{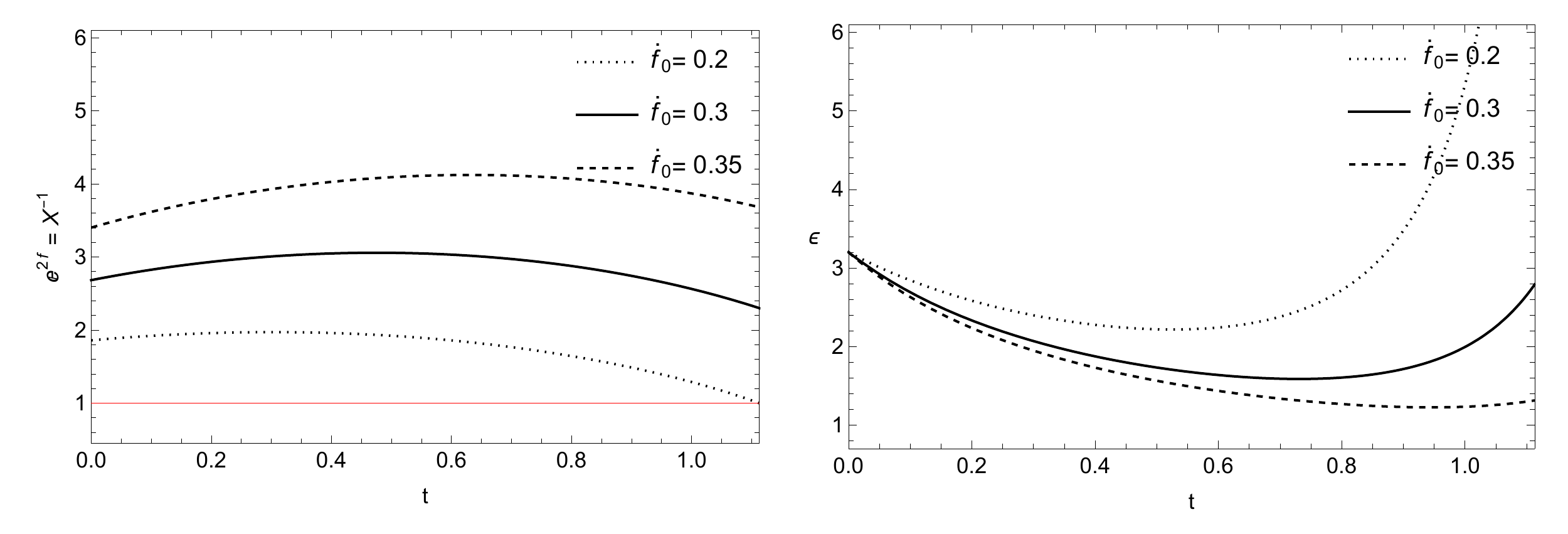}
    \caption{The time evolution for de Sitter universe with $H=0.2$. The \emph{top-left} figure shows the solution flow, where the dotted-, solid-, and dashed-black curves are generated by an initial condition $\{\dot{f}(t_f),\epsilon\}=\{(0.2,0.3,0.35),3.2\}$, respectively. The brown- and red-shaded regions are the same as those in Figure~\ref{fig:01}, while the gray-shaded region represents the parameter space violating the NEC ($\omega_3<-1$). The black dot denotes the stable fixed point. The \emph{top-right} figure presents the scalar-field potential obtained for the dotted-, solid, and dashed-black trajectories in the \emph{top-left} figure. The \emph{bottom-left} figure depicts the convex evolution of the size of extra dimensions, while the \emph{bottom-right} figure shows the bouncing evolution of the energy density. } \label{fig:05}
\end{figure}
To see how the potential can be constructed, let us take the constant $H$ describing de Sitter universe into account. Since $H$ is a constant, $\dot{H}$ is dropped out in the flow equation (\ref{dynamical00}). Together with $\omega_3=\omega_4$ for this case, we obtained the following flow equation:
\begin{align}\label{dSflowEq}
&\ddot{f}_{\text{dS}}=13 \dot{f} H+2 \dot{f}^2+6 H^2+3 X-\epsilon\,,\nonumber\\
&\dot{\epsilon}_{\text{dS}}= 2 \left(4 \dot{f}+3 H\right) \left(24 \dot{f} H+6 \dot{f}^2+12 H^2+6 X-2 \epsilon \right)\,.
\end{align}
The equation of state of this case can be written as
\begin{align}
\omega_3 = 3-\frac{6 }{\epsilon}\left(8 \dot{f} H+2 \dot{f}^2+4 H^2 +2 X\right)\,,\label{Sflow1}
\end{align}
where the inverse size $X$ is given as $X_+$ in (\ref{Inverse S}). Employing (\ref{dSflowEq}), we can generate a solution flow for a certain value of $H$. The solution flow is shown in Figure \ref{fig:05}. Although it is possible to find all fixed points for this solution flow, we only present the stable fixed point in the figure, the black dot at a position of $\{\dot{f},\epsilon\}=\{H,21 H^2\}$. The corresponding inverse size is $X=0$; hence, this fixed point describes a de Sitter space with the infinite size of extra dimensions.  

To construct a potential of the scalar field, we pick an initial condition and find a numerical solution. The corresponding time evolution of the energy density and $X$ can be seen in Figure \ref{fig:05}. This numerical solution can give us (\ref{Sflow1}) as a function of time. Thus the potential shape can be obtained by utilizing (\ref{scalar integration}) and (\ref{K-V}). The resultant shape is also presented in Figure \ref{fig:05}. One can find another expression of the potential using (\ref{Curvature4}) as follows:
\begin{align}
16 \pi  G_8 V= 6 \left(4 \dot{f} H+3 H^2+X^2\right)\,.
\end{align}
The potential shape is sensitive to the choice of $H$, and one may choose a more general form of $H(t)$. Such an extension will be reported soon \cite{Scalar-IUniverse}.

\section{Discussion}\label{Discussion}

In this work, we study the effects of matter on the cosmological evolution driven by an instanton smeared on the extra dimensions. In the previous work \cite{Kim:2018mfv}, we showed that dynamical compactification results in the expanding universe. The corresponding cosmological evolutions or the solution flows can be characterized by two kinds of dynamics. One is the expanding universe with contracting extra dimensions, and the other one is the contracting universe with expanding extra dimensions. In addition, the solution flows have a non-constant expansion or contraction rate.  

On the other hand, the inclusion of matter in the present work can generate various kinds of cosmic evolutions. As a first try, we consider the cosmological constant matter to understand how the solution space changes. Even for this simplest case, various solution structures appear, and the matter generates various fixed points. Analyzing these fixed points, it is possible to classify the full structure of solution flows driven by the cosmological constant. There are interesting time evolutions, such as bouncing universes and a de 
Sitter fixed point. The bouncing behavior is ubiquitous in this model due to the volume-preserving nature of Einstein gravity. In addition, we found a de Sitter solution given by a stable fixed point, as well as a spirally fixed point, according to the value of the cosmological constant. At this fixed point, we evaluate the Hubble parameter in the unit of the 4--dimensional Planck scale regarded as an observed cosmological constant. It turns out that the estimated Hubble constant (\ref{H_p l_p4}) is extremely small even for an 8--dimensional cosmological constant, which is not too small. This can address the cosmological constant problem in this toy universe model. The origin of this small cosmological constant is based on the dynamics of the extra dimensions and the nonlinear nature of the flow equations having a nontrivial fixed point. 

As another try, we considered the matter whose energy density (\ref{energydensity}) depends on the volume of the extra dimensions. In this case, we impose the NEC for the matters as a physical condition, and the typical time evolution is again a bouncing universe. We accomplished the analysis for all the fixed points. The result of the time evolution contains negative energy regions, but flows of these regions still show expanding or contracting behavior of the universe. This can be generalized by a more general form of energy such as $\epsilon=\epsilon(X)$. Then, one has to find the NEC for this case for consistency. It would be desired to see if this kind of generalization can be related to phenomenological models. 

Our final consideration employs the scalar field with potential as a matter in this model, and we focus on the de Sitter universe balanced by potential. This scalar matter case is an extension of a usual cosmological model by inclusion of the extra dimension dynamics, driven by an instanton. A lesson from this study is the potential has a nontrivial profile due to the extra dimension effect. To preserve the de Sitter universe, the energy density shows a bouncing behavior, and the examples introduced here have rolling-type potentials, see Figure \ref{fig:05}. This is an inverse problem that asks what kind of potential generates de Sitter universe. This question can be generalized for a time-dependent Hubble parameter $H(t)$. We will include such a study in our upcoming paper \cite{Scalar-IUniverse}.

Most resultant time evolutions of this model contain $X\geq1$ regions, where the size of extra dimensions is smaller than the Plank scale. These regions appear in the late or early times in the time-dependent geometry. Since the volume size is comparable to the Plank scale, the classical analysis breaks down. Thus we need to consider a quantum gravity effect to describe solutions in these regions. A possible quantum gravity effect of this model is a higher derivative correction such as a Gauss-Bonnet term. A study of this correction can give us an insight into the final configuration of the dynamical compactification. We also leave this as a future study. {\color{black} As is mentioned, the analyses of the present study are focused on the Jordan frame. Therefore, the bouncing solutions obtained in this work may not be translated into the bouncing solutions in the 4--dimensional Einstein frame. Thus, it is worth investigating the solutions in the Einstein frame and analyzing them to show whether bouncing without the violation of the NEC is possible in the Einstein frame.  }

We introduce the cosmological constant matter to understand how the solution structure changes by matter. If we consider more extra dimensions than the 8--dimensions with a flux wrapping these newly introduced dimensions, then such an additional structure can generate an effective cosmological constant to the 8--dimensional spacetime. Furthermore, this construction could be the origin of the non-abelian gauge fields. It would be interesting to realize this toy model in 10-- or 11--dimensions by introducing 2 or 3 extra compact manifolds which have a non-abelian structure. Also, a study of cosmological perturbation on this background is an important subject. We hope to find the cosmological perturbation on the backgrounds we obtained here. 

\appendix

\section{Decomposition of Einstein Equation and 4--dimensional actions}\label{Decom-Einstein}

The decomposition of the Einstein equation based on (\ref{metric00}) is given by
\begin{align}\label{EinsteinGEq}
R^{(g)}_{\mu\nu} -\frac{1}{2}R^{(g)}g_{\mu\nu} &=\, -\left(4\nabla^{{(g)}2}f + 10 \nabla^{(g)}f\cdot \nabla^{(g)}f - \frac{1}{2}e^{-2f}R^{(h)} \right)g_{\mu\nu}\nonumber\\
&~~+4 \left(\nabla^{(g)}_\mu\nabla^{(g)}_\nu f+\partial_\mu f\partial_\nu f\right)-\frac{8\pi G_{(8)}}{g_{\text{YM}}^2}e^{-4f}\rho_1 g_{\mu\nu} +8\pi G_{(8)} T^m_{\mu\nu}\,,\nonumber\\
R^{(h)}_{ab}-\frac{1}{2} R^{(h)} h_{ab}&=\, 3 e^{2f(x)}\left(\frac{1}{6}R^{(g)}-\nabla^{(g)2}f - 2\nabla^{(g)}f \cdot \nabla^{(g)}f \right)h_{ab} + 8\pi G_{(8)} T^m_{ab}\,.
\end{align}
Here, one can see that $\rho_1$ should be a constant to satisfy the Einstein equation with the perfect fluid matter whose energy density and pressures have only time dependence.

{\color{black}These equations of motion without matter can be reproduced by a 4--dimensional action, which is obtained by a consistent truncation using $\mathbb{S}^4$ compactification. The effective 4--dimensional action is
\begin{align}
\mathcal{I} = \frac{\mathcal{V}_4}{16\pi G_{(8)}}\int d^4x \sqrt{-g}\left(e^{4f} R^{(g)}+ 12\, g^{\mu\nu}\partial_\mu f\partial_\nu f + \mathcal{A} e^{2f}-\mathcal{B}\right)\,,
\end{align}
where $\mathcal{A}=\frac{48\pi}{\sqrt{6}}\frac{1}{\sqrt{\mathcal{V}_4}}$ and $\mathcal{B}=\frac{16\pi G_{(8)}}{\mathcal{V}_4}\frac{8\pi^2}{g_{\text{YM}}^2}$. Also, one may express this action in the 4-dimensional Einstein frame. The expression is given by
\begin{align}
\mathcal{I} = \frac{\mathcal{V}_4}{16\pi G_{(8)}}\int d^4x \sqrt{-\tilde{g}} \left(\tilde{R} - 12\, \tilde{g}^{\mu\nu} \partial_\mu f \partial_\nu f + \mathcal{A}e^{-6f} - \mathcal{B} e^{-8f} \right)\,.
\end{align}
The above two actions are convenient to deal with our model. When, however, additional matters are introduced to the original system, one should be very careful. The origin of these actions comes from the 8--dimensional theory, so the physical energy conditions and the energy-momentum tensor must be considered from the 8--dimensional point of view.
}

\section{Another type of flow equation}\label{AnotherType}

Let us consider (\ref{flowEq0}) again to find another slicing of the solution space. We will express the equations in terms of $S(t)=e^{2f}$, $\dot{S}$ and $\ddot{S}$ rather than $e^{-2f}$, $\dot{f}$ and $\ddot{f}$. The alternative equations can be written as follows:
\begin{align}\label{Flow-S}
&\hat{E}_1=S^2 \left(6 H^2-2 \epsilon \right)+12 S \left(1+H \dot{S}\right)+3 \left(\dot{S}^2-2\right)=0\,,\nonumber\\
&\hat{E}_2=S^2 \left(3 H^2+2 \dot{H}+\omega _3 \epsilon \right)+2 S \left(2 H \dot{S}+\ddot{S}+3\right)+3 \left(\dot{S}^2-1\right)=0\,,\nonumber\\
&\hat{E}_3=12 H^2 S+9 H \dot{S}+6 \dot{H} S+3 \ddot{S}+2 S \omega _4 \epsilon +6=0\,,\nonumber\\
&\hat{E}_4=3 H S \left(\omega _3+1\right) \epsilon +2 \dot{S} \left(\omega _4+1\right) \epsilon +S \dot{\epsilon }=0\,.
\end{align}
As one can see, the first equation $\hat{E}_1=0$ is the constraint equation of the system. Likewise the previous analysis, we may solve the equation in terms of $S$. Then, the size of the extra dimensions is given by
\begin{align}
e^{2f}=S_\pm=\frac{-6\, (1+H \dot{S})\mp\sqrt{6} \sqrt{3 H^2 \left(\dot{S}^2+2\right)+12 H \dot{S}+\left(\dot{S}^2-2\right) \epsilon +6}}{6 H^2-2 \epsilon }\,.
\end{align}
This equation gives two solution sets again. One is the set with $S(t)=e^{2f}\geq -3\, (1+H \dot{S})/(3 H^2- \epsilon)$ and the other is the set with the extra dimension size function $S(t)\leq-3\, (1+H \dot{S})/(3 H^2- \epsilon)$. One may notice that this boundary of the solution space is not $S=1$. These equations of motion (\ref{Flow-S}) can be regarded as a different slicing of the solution set by the constraint equation $\tilde{E}_1=0$, which is different from that of (\ref{flowEq0}). This set of equations \ref{Flow-S} can describe how solutions evolve near $S=1$. However, solutions with $S(t)$ smaller than 1 have no meaning as classical solutions due to the strong quantum gravity effect. 

\section*{Acknowledgments}

We thank Hyun Seok Yang, Seokcheon Lee, Miok Park, and Yunseok Seo for their helpful discussions. This work is supported by Mid-career Research Program through NRF grant No. NRF-2019R1A2C1007396 (K. Kim) and No. NRF-2021R1A2C1005748 (S. Koh and G. Tumurtushaa). This work is also supported by Basic Research Program through NRF grant No.  2022R1I1A1A01053784 (G. Tumurtushaa).

\providecommand{\href}[2]{#2}\begingroup\raggedright

\endgroup

\providecommand{\href}[2]{#2}\begingroup\raggedright
\bibliography{references}
\bibliographystyle{JHEP}
\endgroup

\end{document}